\documentclass[prb,aps,twocolumn,amsmath,amssymb,floatfix,
superscriptaddress]{revtex4-2}
\usepackage[dvips]{graphics}
\usepackage[justification=raggedright]{caption}
\usepackage{subcaption}
\usepackage[font=footnotesize,figurewithin=none]{caption}
\usepackage{subcaption}
\usepackage{adjustbox}
\usepackage[labelfont=bf]{caption}
\usepackage[labelformat=parens, labelfont=bf]{subcaption}
\usepackage{color}
\usepackage{float}
\usepackage{natbib}
\usepackage[dvipsnames]{xcolor}
\setcitestyle{numbers}
\setcitestyle{square}
\definecolor{dred}{rgb}{0,0,0.6}
\usepackage{graphicx}    
\usepackage{bm}  
\usepackage{amssymb} 
\usepackage{mathtools}
\usepackage{mathdots}
\usepackage{amsmath}
\usepackage{braket}
\usepackage[mathscr]{euscript}
\usepackage[colorlinks=true, allcolors=blue]{hyperref}
\usepackage{url}
\begin{document}
\title{Topological properties of a periodically driven Creutz ladder}

\author{Koustav Roy and Saurabh Basu \\ \textit{Department of Physics, Indian Institute of Technology Guwahati-Guwahati, 781039 Assam, India}}

\date{\today}
\begin{abstract}
We have investigated a periodically driven Creutz ladder in presence of two different driving protocols, namely, a sinusoidal drive and a $\delta$-kick imparted to the ladder at regular intervals of time. Specifically, we have studied the topological properties corresponding to the trivial and the non-trivial limits of the static (undriven) case via computing suitable topological invariants. Corresponding to the case where the chiral symmetry of the ladder is intact, in addition to the zero energy modes, $\pi$ energy modes appear in both these cases. Further, two different frequency regimes of the driving protocol emerge, where the Floquet-Magnus expansion is employed particularly to study the high frequency regime for the sinusoidal drive. Apart from the physics being identical in the high frequency and the static scenarios, the zero energy modes show distinctive features at low and high frequencies. For the sinusoidal drive, there exists a sharp frequency threshold beyond which the zero energy modes only exist in the topological limit, while in the trivial limit, it exist only upto the same threshold frequency. In presence of the $\delta$-kick, the Creutz ladder demonstrates higher values of the topological invariant, and as a consequence, the system possesses larger number of edge modes.
\end{abstract}

\maketitle

\begin{center}\section{\label{sec:level1}Introduction}\end{center}
\par Topological insulators are materials that exhibit bulk states which are gapped, similar to a conventional insulator, whereas the states at the boundaries are gapless. These states contribute to the electronic transport properties. The presence of these localised boundary states is solely determined by the bulk properties of the system (`bulk-edge correspondence'), and are immune to local perturbations that do not close the spectral gap and hence allow for non-dissipative electronic transport \cite{L.zhang,Cooper2019,hugelchiral,jintopological,jinbulk}. Creutz ladder is an example of such a material having quasi-1D structure and consists of two rungs of lattice sites that are coupled by diagonal, vertical and horizontal hoppings. Additionally, a magnetic flux can penetrate the ladder perpendicular to its plane, giving access to an extra degree of freedom in the form of Peierls phase \cite{cruetz1999,Gholizadehcreutz,mukherjeetailoring}, associated with the horizontal hopping amplitude. Moreover, the localisation of the zero energy modes are also characterised in terms of Aharonov-Bohm caging. Due to this dual protection (Aharonov-Bohm caging and topological symmetry conservation), the edge modes are very robust irrespective of the system size. These intriguing properties may make Creutz ladder a promising candidate for different applications in high performance electronics and quantum computing.\\

The dynamics of quantum systems is another topic of considerable interest, and is actively explored in the field of topology. Recent studies have established that applying external periodic drives can open a route to engineer topological non-triviality with high tunability from materials that are even topologically trivial in equilibrium \cite{cayssolwiley,rudner2013,gomezfloquet,rudner2020}. Furthermore, because of the periodicity in time, the energy bands are folded back to a Floquet Brillouin zone (FBZ), at the boundary of which a new variety of edge modes, namely, the so called $\pi$ modes appear.
The value of the topological invariants, predicting the number of the zero and the $\pi$ quasi-energy end states can be controlled by the parameters, such as, frequency and amplitude of the drive. In general, the Floquet topological insulators (FTIs) show rich topological phases that may not have any analogy with the undriven case. For example,  generation of higher Chern number in 1D extended Su-Schrieffer-Heeger (E-SSH) model \cite{agrawalssh,wangssh}, emergence of time crystalline phase and period doubling oscillations in 1D time Floquet SSH \cite{period2t,wangperiod2t}, rich entanglement properties of the time periodic Kitaev chain \cite{yateskitaev,mondalkitaev}, Floquet analysis of higher order topological insulators \cite{ghoshfloquet,seshadri,klinovaja}. These works have continued to draw attention from the community owing to the experimental success in driving quantum systems. For example, the experimental realization of FTI is done on a nano-photonics platform using a lattice of strongly coupled octagonal resonators in the silicon-on-insulator material system \cite{afzal}. Further, the time periodicity of any system may be exploited by the means of photo induced band gaps, which further can be resolved by the technique called time and angle resolved photoemission spectroscopy (t-ARPES) \cite{wangarpes,leearpes} .
\\

The applications of Floquet dynamics is well explored in the field of 1D topological insulators, like SSH chain or Kitaev chain with $p$-wave superconductivity \cite{lago,senkitaev,yuoptical,senkitaev2}. But it has not been explored whether we can get such rich topological properties from quasi-1D system, just like Creutz ladder. In this work, we shall highlight the Floquet aspects captured by the ladder when the hopping amplitudes are modulated in  time by two different driving protocols, namely, (a) a sinusoidal drive and (b) a $\delta$-kick. We shall also provide a comparison between the two protocols and the corresponding insights induced by them. In a generic sense, these two drives account for any kind of periodically driven systems. In case of a $\delta$-kick, the Hamiltonian at different times commute which makes the time evolution trivial, and hence we can easily get access to the stroboscopic Hamiltonian. Whereas, in the case of a sinusoidal drive, the Hamiltonian at different times do not commute, and hence the time evolution becomes non-trivial. However, the time dependence of the Hamiltonian for the sinusoidal drive can be eliminated by a similarity transformation, following which the factorization of the evolution operator is nothing but a rotating frame transformation followed by an expansion in powers of the inverse of the driving frequency, namely, $1/\omega$ (Floquet-Magnus expansion) \cite{eckardt,wangmultiple,zengpwave}. Our primary goal is here to compare and contrast time independent Floquet Hamiltonians, constructed from these two drives.\\

\par The layout of the subsequent discussion is as follows. In section II, we describe the static (undriven) version of the model to recapitulate its symmetries. We also introduce the Floquet formalism and briefly discuss the spectra. In section III, we shall discuss our results on the sinusoidal drive, where the explicit time dependence is eliminated via Shirley-Floquet approach. Further, we have explored the analytical behaviour of the Hamiltonian in the high frequency regime. Finally, we shall discuss the $\delta$-kick scenario, where easy factorization of the time evolution operator yields a way to analyse various topological invariants. At the end, we summarize and conclude our findings in section IV.
\begin{figure}[!h]
    \begin{subfigure}[b]{\columnwidth}
         \includegraphics[width=\columnwidth]{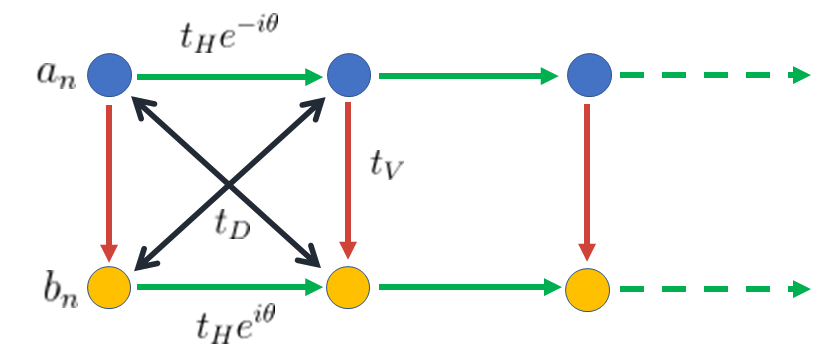}
         \label{1}
     \end{subfigure}
\caption{{The figure depicts a schematic representation of the quasi 1-D Creutz ladder, where $a_n$ and $b_n$ denote the two distinct sublattices. The different hopping amplitudes, $t_H$, $t_V$, $t_D$ denote the horizontal, vertical and diagonal hoppings respectively.}} 
\label{1}
\end{figure}
\vspace*{-0.45cm}
\begin{center}{\section{\label{sec:level2}The Hamiltonian and the Floquet Formalism}}\end{center}
\begin{figure}[!h]
          \begin{subfigure}[!h]{0.493\columnwidth}
         \includegraphics[height=45mm,width=\columnwidth]{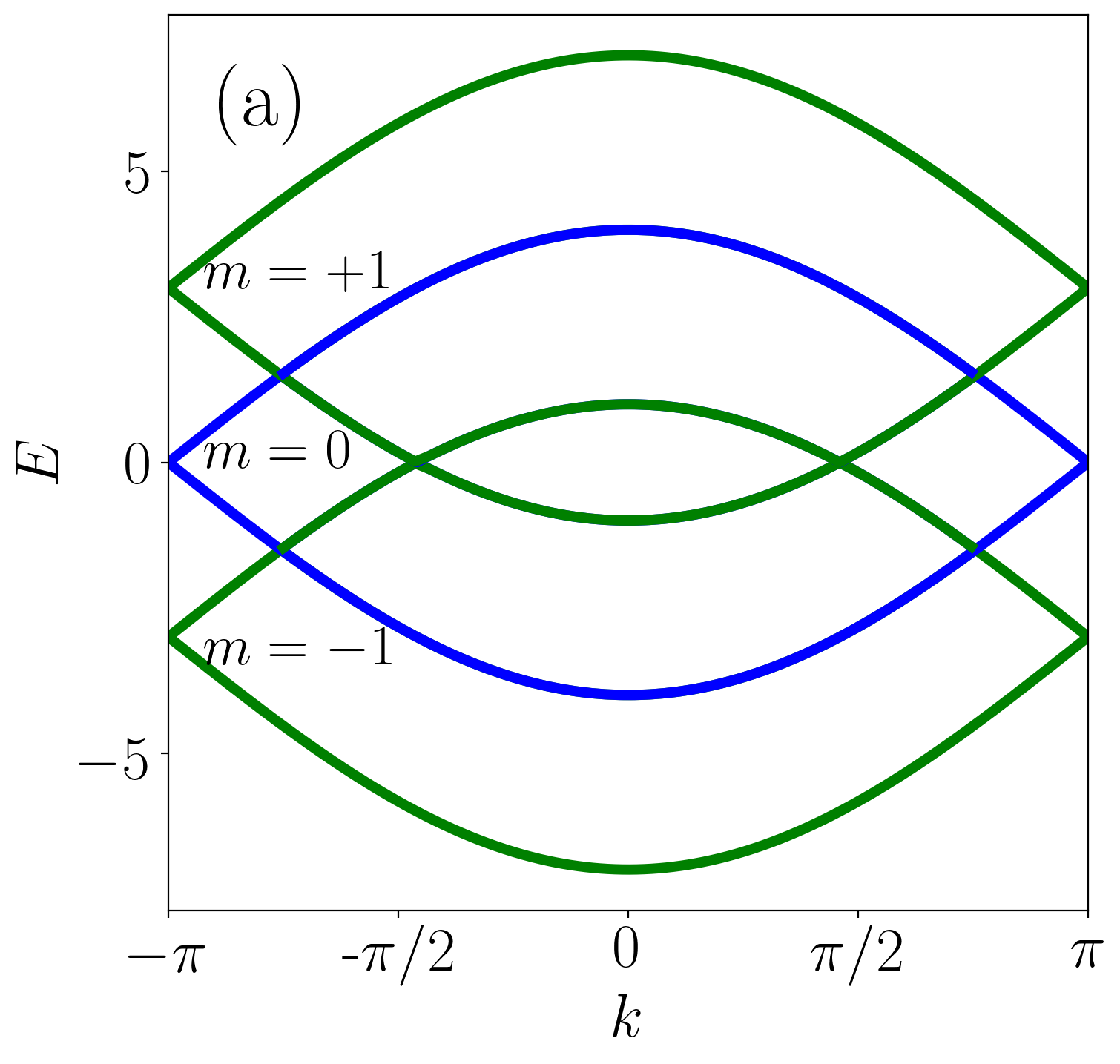}
         \captionlistentry{}
         \label{2.1}
     \end{subfigure}
     \begin{subfigure}[!h]{0.493\columnwidth}
         \includegraphics[height=45mm,width=\columnwidth]{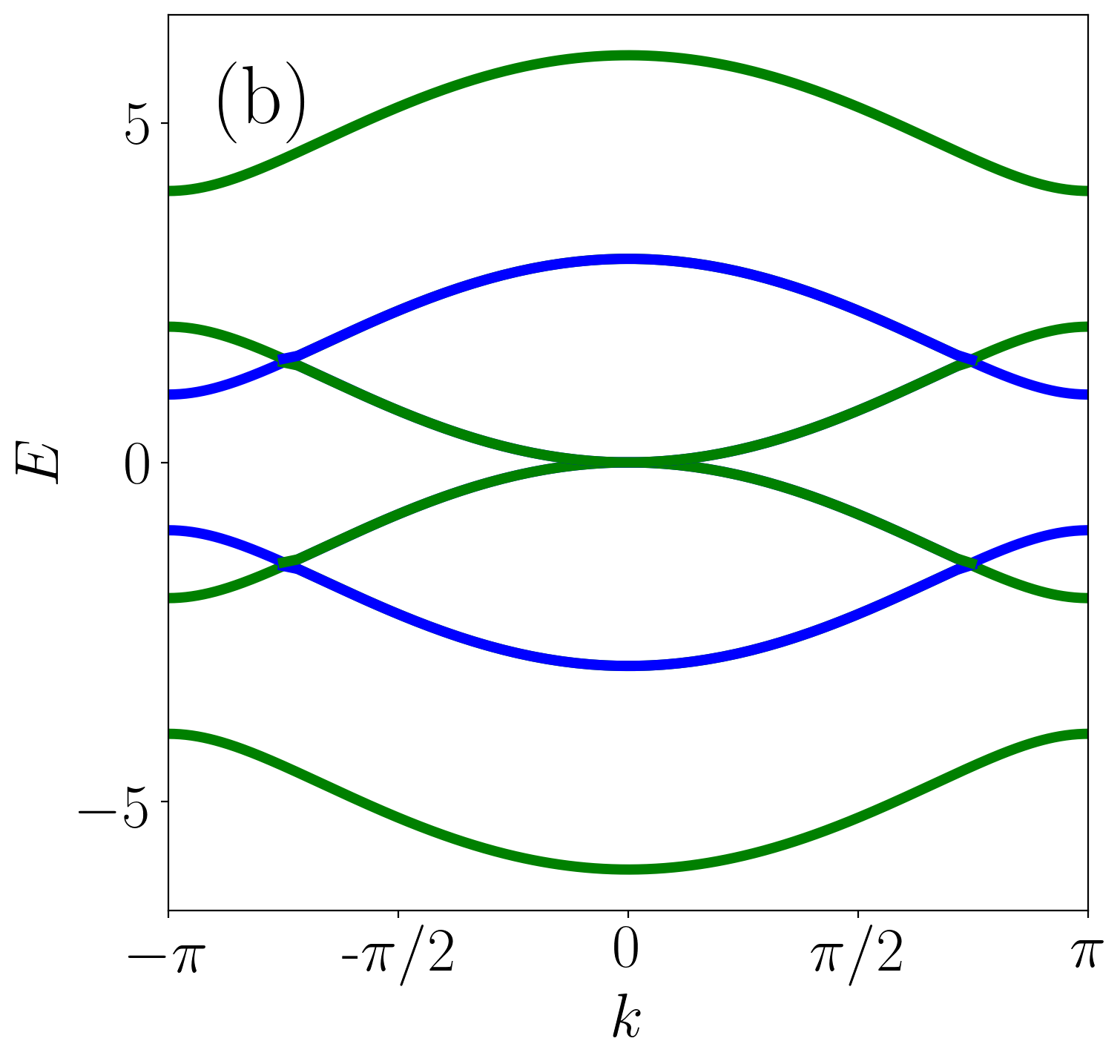}
         \captionlistentry{}
         \label{2.2}
     \end{subfigure}
     \begin{subfigure}[!h]{0.493\columnwidth}
         \includegraphics[height=45mm,width=\columnwidth]{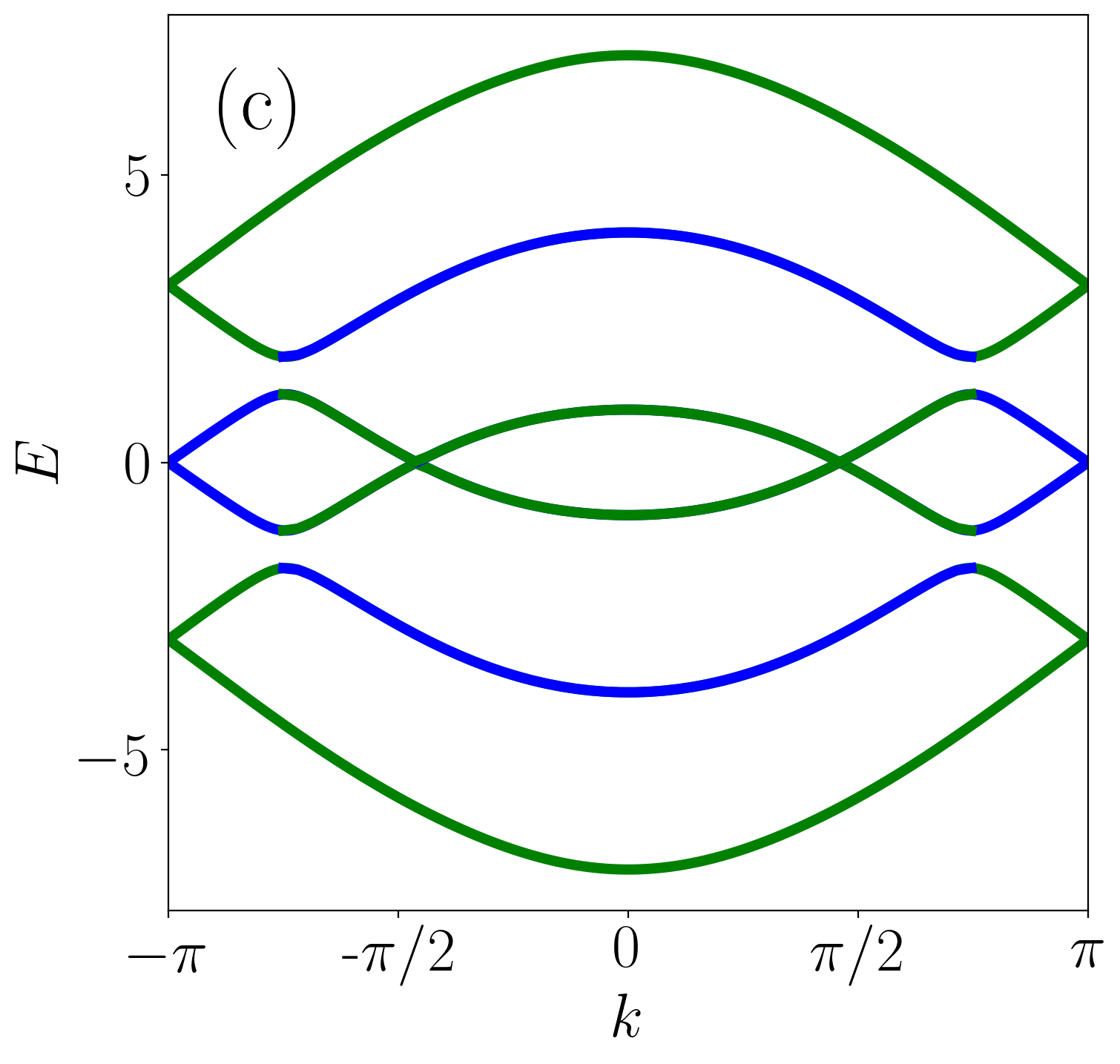}
         \captionlistentry{}
         \label{2.3}
     \end{subfigure}
     \begin{subfigure}[!h]{0.493\columnwidth}
         \includegraphics[height=45mm,width=\columnwidth]{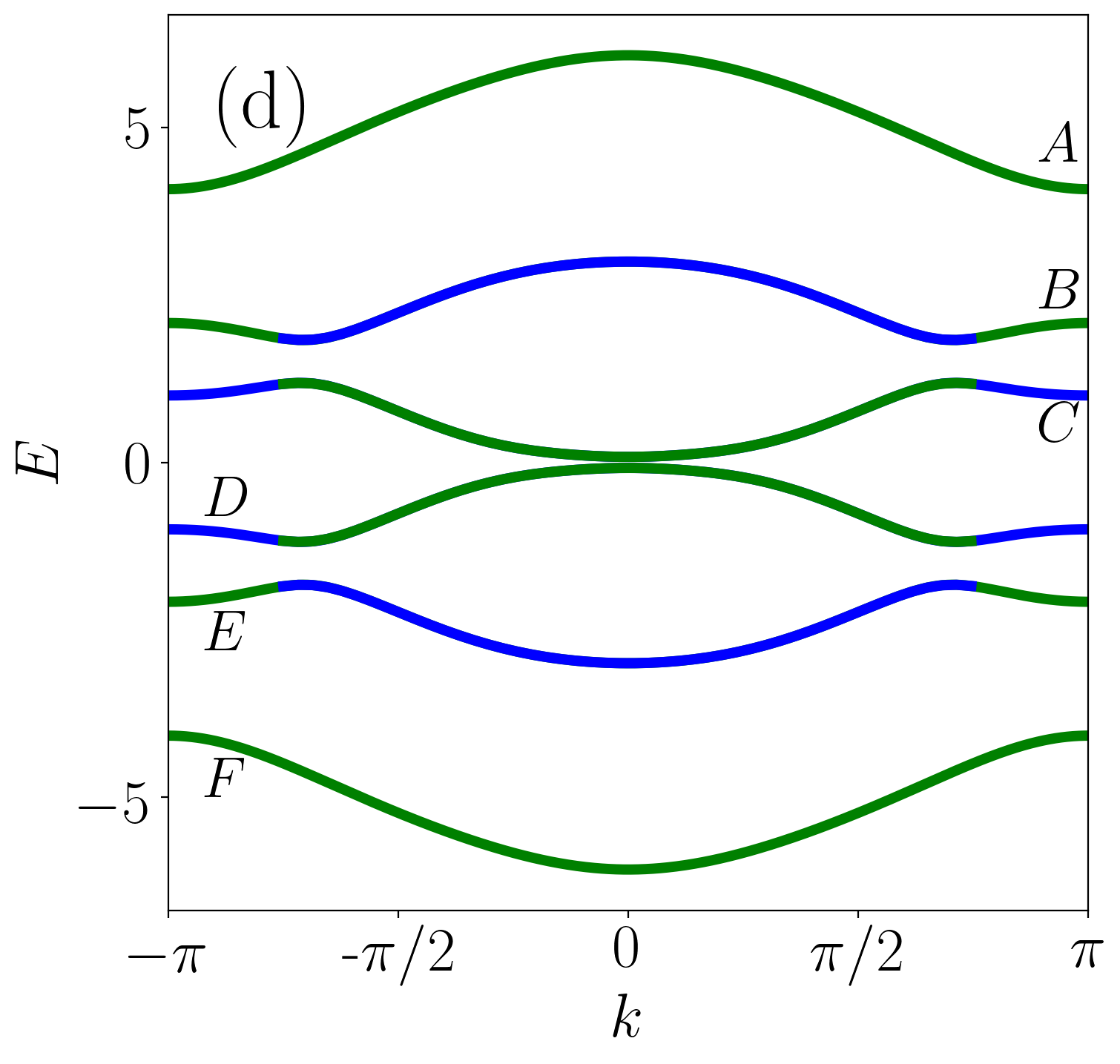}
         \captionlistentry{}
         \label{2.4}
     \end{subfigure}
\caption{{The Floquet quasi-energy spectrum in the frequency space with $m \in [-1,1]$, obtained for different sets of values of the hopping parameters as well as the driving amplitude, $V_0$. The opening and closing of the zero and the $\pi$ gaps are clearly illustrated. The parameters used are, $(t_V, t_D) = (2, 1)$ in panel (a) and (c) and $(t_V, t_D) = (1, 1)$ in panel (b) and (d). The amplitude of the drive is chosen as, $V_0=0$, for panel (a) and (b), whereas $V_0 = 0.5$, for panel (c) and (d). The rest of the parameters are chosen as, $t_H=1,\omega=3,\theta=\pi/2$.}} 
 \label{2}
\end{figure}
\par The Creutz ladder consists of two rungs of lattice sites that are coupled by diagonal ($t_D$), vertical ($t_V$) and horizontal ($t_H$) hoppings, as shown in Fig.\ref{1}. There are two sublattices $a_n$, $b_n$ within each unit cell. The real space Hamiltonian can be written as,
\begin{equation}
\label{Eq1}
\begin{split}
&H_0 = -\sum_n t_{H}(e^{i\theta}a_{n}^{\dagger}a_{n+1} + e^{-i\theta}a_{n+1}^{\dagger}a_{n} +e^{-i\theta}b_{n}^{\dagger}b_{n+1} \\& \qquad+e^{i\theta}b_{n+1}^{\dagger}b_{n}) +t_{D}(a_{n}^{\dagger}b_{n+1}+b_{n+1}^{\dagger}a_{n}+a_{n+1}^{\dagger}b_n\\&\qquad+b_{n}^{\dagger}a_{n+1})+t_{V}(a_n^{\dagger}b_n+b_n^{\dagger}a_n),
\end{split}
\end{equation}
\begin{equation}
    H_0 = H_H + H_D + H_V.
\end{equation}
The complex phase, $\theta$ associated with the horizontal hopping leads to a destructive interference, as a consequence of which localisation of particles for a certain region of parameter space is observed. The Creutz ladder shows a flat band dispersion for the rungless case ($t_V=0$), implying that group velocities of the resulting states are zero. In an open boundary condition, this leads to complete localisation of states at the edges.\\

\par In momentum space, the Hamiltonian reads as,
\begin{equation}
\begin{split}
H_{0}(k)=2t_{H}\cos(k)\cos(\theta)\sigma_0&+2t_{H}\sin(k)\sin(\theta)\sigma_z\\&+(t_V+2t_{D}\cos(k))\sigma_x.
\end{split}
\end{equation}
Here $\sigma_i=x,y,z$ are the Pauli matrices. If $\phi$ denotes the total flux through each plaquette then, $2\theta=\frac{\phi}{\phi_0}$, where $\phi_0$ denotes the magnetic flux quantum.\\

\par At this point it is important to talk about the symmetries of the model \cite{licharacterization,hughes,jafari,juneman}. The model has an inherent inversion symmetry with respect to an axis that lies symmetrically between the two legs of the ladder. It is expressed by the relation $\sigma_x H_{0}(k)\sigma_x=H_{0}(-k)$. Furthermore, it possesses a chiral symmetry that is illustrated by, $\sigma_y H_{0}(k)\sigma_y=-H_{0}(k)$, only for the values $\theta=\frac{\pi}{2}$. Inspite of the presence of a magnetic field, the system has an inherent time reversal symmetry given by, $\sigma_x H^{*}_{0}(k)\sigma_x=H_{0}(-k)$. Lastly, a particle-hole symmetry exists in the system for $\theta=\frac{\pi}{2}$ which can be illustrated by, $\sigma_z H^{*}_{0}(k)\sigma_z=-H_{0}(-k)$.\\

\par Now we are all set to study the Floquet topological aspects of the periodically driven model, which in a Creutz ladder enters through the modulation of the hopping amplitudes in time. To set the notations, let us start our discussion for a generic Hamiltonian, by considering the driving in form of a sinusoidal variation, that is, $H(t)= H_0 + 2V_{0}\cos{\omega t}$, with $V_0$ and $\omega$, being the driving amplitude and the driving frequency respectively and $H_0$ is a generic static Hamiltonian. The properties of interest in our work can be obtained using Floquet theory \cite{hanggi,goldmanfloquet,restrepo}, according to which, the time-dependent Schr\"{o}dinger equation (TDSE) can be solved using the Floquet ansatz, $\ket{\psi(t)}=e^{-i E t}\ket{u(t)}$, where $\ket{u(t+T)}=\ket{u(t)}$ denotes the time periodic Floquet modes, and $E$ represents the Floquet quasi-energies.
These Floquet states are also the eigenstates of the Floquet evolution operator. We can find $E$ and $\ket{u_k(t)}$ by solving the Floquet-Bloch equation,
\begin{equation}
[H(t)-i\partial_t]\ket{u_k(t)}= E \ket{u_k(t)}.
\end{equation}
The operator $H(t)-i\partial_t = H_F$ is termed as Floquet Hamiltonian. Because of the time periodicity, it is convenient
to consider the composite Hilbert space $\mathscr{R}\otimes \mathscr{T}$ where $\mathscr{R}$ is the usual Hilbert space with a complete set of orthogonal basis, and $\mathscr{T}$ is the space of time periodic functions spanned by $e^{-im\omega t}$. This yields the following form of $H_F$,
\begin{equation}
H_F=\sum_{m,m^\prime} \Big( m\omega \delta_{m,m^{\prime}} + \frac{1}{T} \int_{0}^{T} dt H(t) e^{-i(m-m^\prime)\omega t} \Big).
\end{equation} 
This leads to a situation where we can split the driven spectrum into an infinite number of copies of the undriven Hamiltonian separated by $m\omega$ that is, the index $m$ defines a subspace, called as the $m^{th}$ Floquet replica. A general representation of the Floquet Hamiltonian thus can be represented as,
\begin{equation}
\label{E6}
    H_{F} = \scriptsize{\begin{bmatrix}\ddots & \vdots  & \vdots & \vdots &\vdots & \vdots & \iddots \\ \dots & H_0 - 2\omega & H_{-1} & H_{-2} & H_{-3} & H_{-4} & \dots \\ \dots & H_1 & H_0  - \omega & H_{-1} & H_{-2} & H_{-3} & \dots \\ \dots & H_2 & H_1 & H_0 & H_{-1} & H_{-2} & \dots \\ \dots & H_3 & H_2 & H_1 & H_0 +\omega & H_{-1} & \dots \\
    \dots & H_4 & H_3 & H_2 & H_1 & H_0 + 2\omega & \dots \\ \iddots & \vdots & \vdots & \vdots & \vdots & \vdots & \ddots
    \end{bmatrix}},
\end{equation}
where the elements $H_{\pm m}=\frac{1}{T}\int^T_0 {H(t)e^{\pm i m \omega t} dt}$ get rid of the explicit time dependence.\\

\par Coming back to our context of Creutz ladder, it is important to understand the consequences of adding periodic drives to either one, two or all of the three hoppings, namely, $t_V,t_D$ and $t_H$. For instance if we include a drive to the horizontal hopping ($t_H$) then in the Fourier space of $H_F$, $H_{\pm m}$ will appear in the form of an identity matrix.
Physically this suggests that the drive fails to induce overlap of different Floquet replicas, and hence the system can not generate the so called $\pi$ energy modes at the boundaries of the Floquet BZ. A clearer picture will emerge from the subsequent discussions. In order to realise meaningful Floquet topological features, we must associate the drive to either the diagonal hopping ($t_D$) or the vertical hopping ($t_V$). In both these cases, the drive connects different sublattice degrees of freedom.
While we have verified that the scenarios for $t_D$ and $t_V$ are qualitatively similar, we have chosen the Floquet driving in $t_V$ for our numeric computation.
\vspace*{1cm}
\begin{center}{\section{\label{sec:level3}Results}}\end{center}
\vspace*{-1.35cm}
\begin{center}{\subsection{\label{sec:level3.1}Sinusoidal drive}}\end{center}
\vspace*{0.1cm}
\par We first describe the harmonic drive, associated only with the vertical hopping ($t_V$), which can be written as,
\begin{equation}
H_{V}(t)=(2V_{0}\cos{\omega t} + t_V)\sum_{n}(a_n^{\dagger} b_n+a_n b_n^{\dagger}).
\end{equation}
The rest of the terms in Eq.\ref{Eq1} are left unaltered. The Fourier components $|H|_{\pm m}$ except for $m=0,\pm1$ vanish owing to the mathematical form of the drive. Hence we can truncate the infinite dimensional matrix into a $3\times3$ block and can study the corresponding quasi-energy spectrum. By using Floquet theory we can show that driving induces additional gaps and edge states depending upon the driving frequency and the strength of the driving field. Before going into that, we re-emphasize that for our purpose the chiral symmetry of the model is of prime importance. Since the emergence of  the $\pi$ energy modes are protected by the chiral symmetry, we fix a particular value of the phase, namely $\theta=\frac{\pi}{2}$. Let us first briefly recapitulate the topological properties of the static model. The topological phase transition is signaled by the ratio $\frac{t_V}{2t_D}$. The system shows non-trivial behaviour for $\frac{t_V}{2t_D}<1$ and trivial behaviour $\frac{t_V}{2t_D}>1$. Fig.\ref{2.1} and Fig.\ref{2.2} show the spectra for $t_V=2t_D$ and $t_V\neq2t_D$ respectively. While the former denotes a gap closing scenario, at $k=\pm \pi$ (The lattice constant is taken as unity), the latter does have a gap at the edges of the BZ with a magnitude $\Delta=2|t_V-2t_D|$. The colour scale
represents different replicas, for example, blue denotes $m=0$, whereas green denotes $m=\pm 1$. The importance of topology can be explored at the points $E=0$,$\pm\frac{\pi}{T}$ ($\pm\frac{\pi}{T}$ is equivalent to $\pm\frac{\omega}{2})$, where the spectra show degeneracies.\\
\begin{widetext}
    \begin{minipage}{\linewidth}
        \begin{figure}[H]
        \centering
        \hspace{-7mm}
          \begin{subfigure}[!h]{0.5\columnwidth}
         \includegraphics[width=\columnwidth]{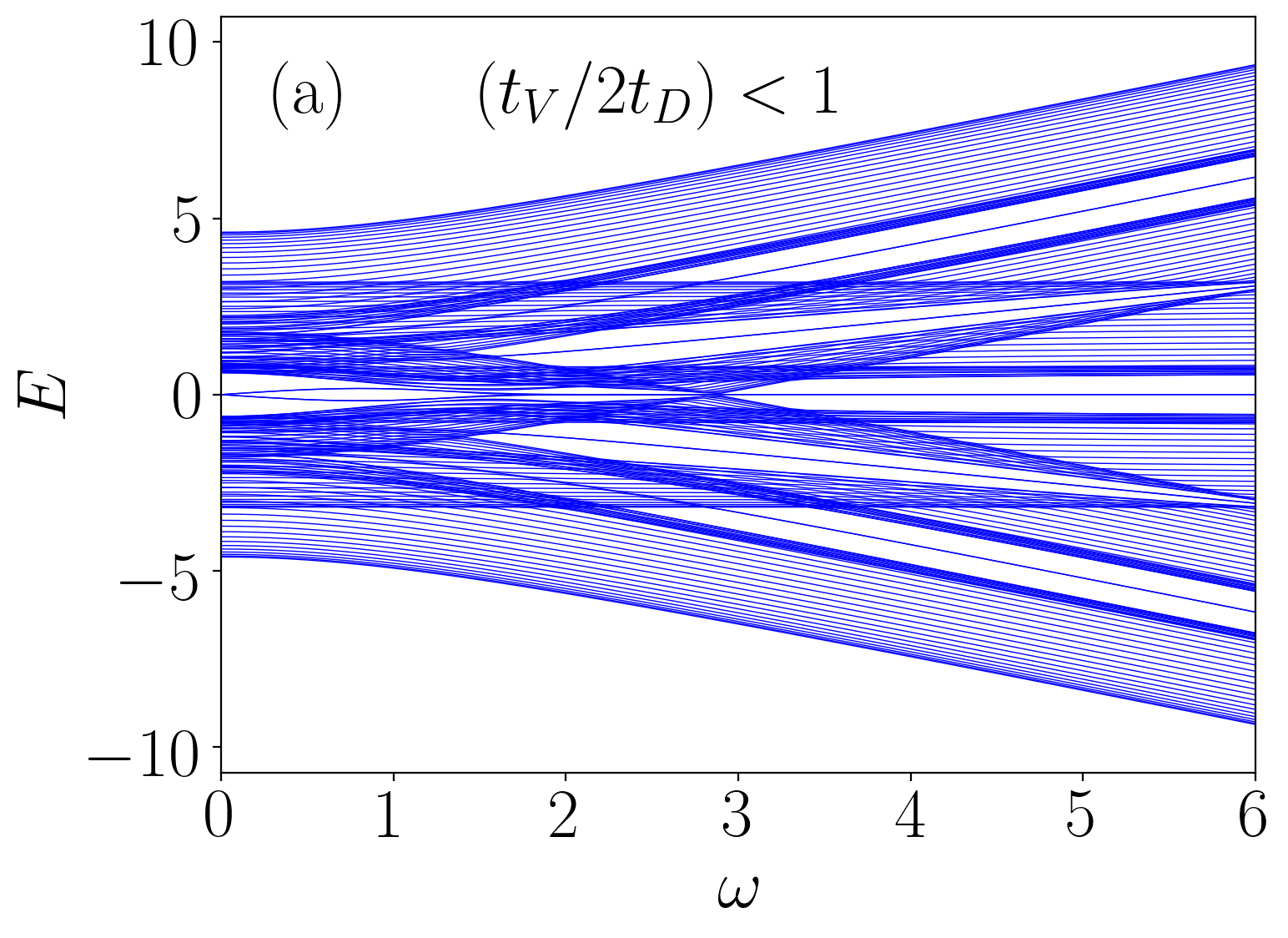}
         \captionlistentry{}
         \label{3.1}
     \end{subfigure}
     \begin{subfigure}[!h]{0.5\columnwidth}
         \includegraphics[width=\columnwidth]{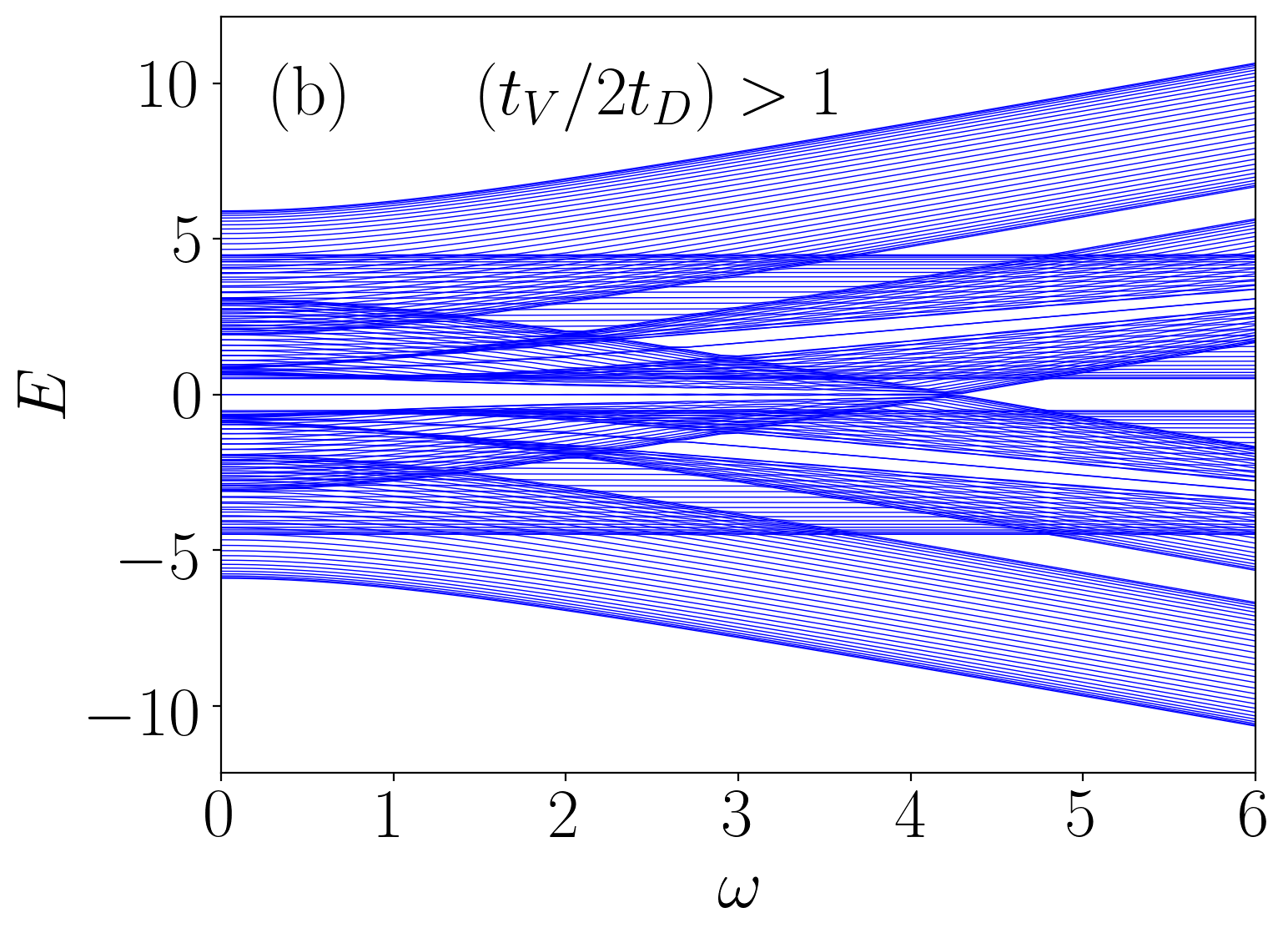}
         \captionlistentry{}
         \label{3.2}
     \end{subfigure}
\caption{{ The Floquet real space quasi-energy spectrum as a function of the driving frequency, $\omega$. Panel (a) corresponds to the static non-trivial condition with the parameter choice, ($t_V, t_D$) = ($1, 1$). Panel (b) corresponds to the static trivial condition with the parameter choice, ($t_V, t_D$) = ($2.2, 1$).
The remaining parameters are chosen as, $t_H=1, \theta = \pi/2, V_0 = 0.5$.}} 
 \label{3}
\end{figure}
        \begin{figure}[H]
          \begin{subfigure}[!h]{0.5\columnwidth}
         \includegraphics[width=1.\columnwidth]{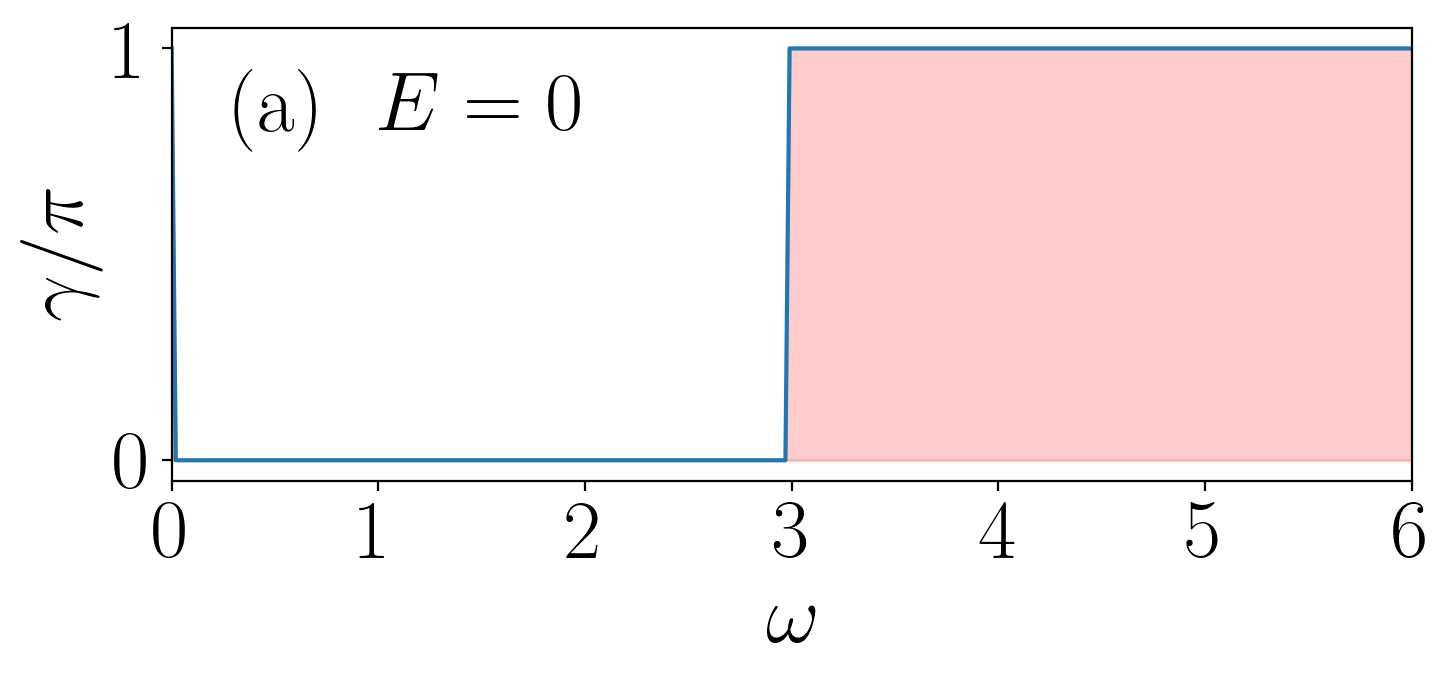}
         \captionlistentry{}
         \label{4.1}
     \end{subfigure}
     \begin{subfigure}[!h]{0.5\columnwidth}
         \includegraphics[width=\columnwidth]{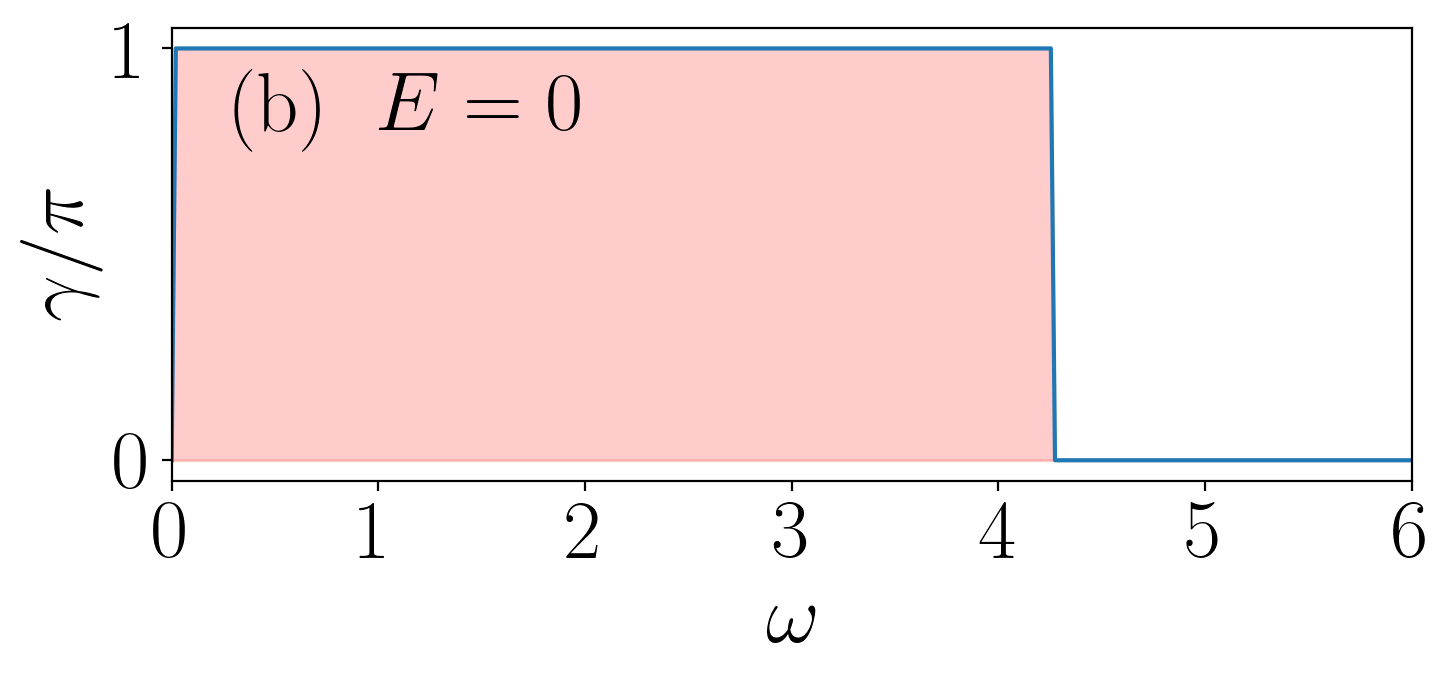}
         \captionlistentry{}
         \label{4.2}
     \end{subfigure}
     \begin{subfigure}[!h]{0.5\columnwidth}
         \includegraphics[width=\columnwidth]{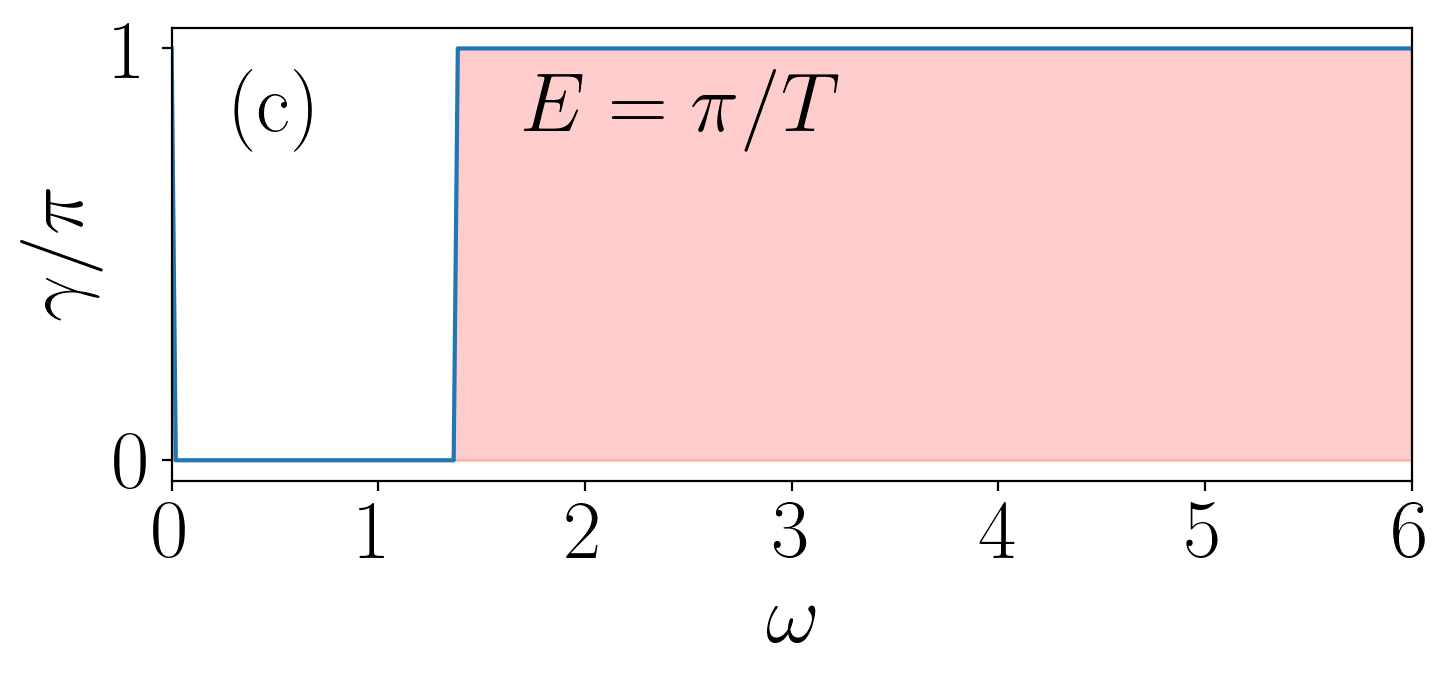}
         \captionlistentry{}
         \label{4.3}
     \end{subfigure}
     \begin{subfigure}[!h]{0.5\columnwidth}
         \includegraphics[width=\columnwidth]{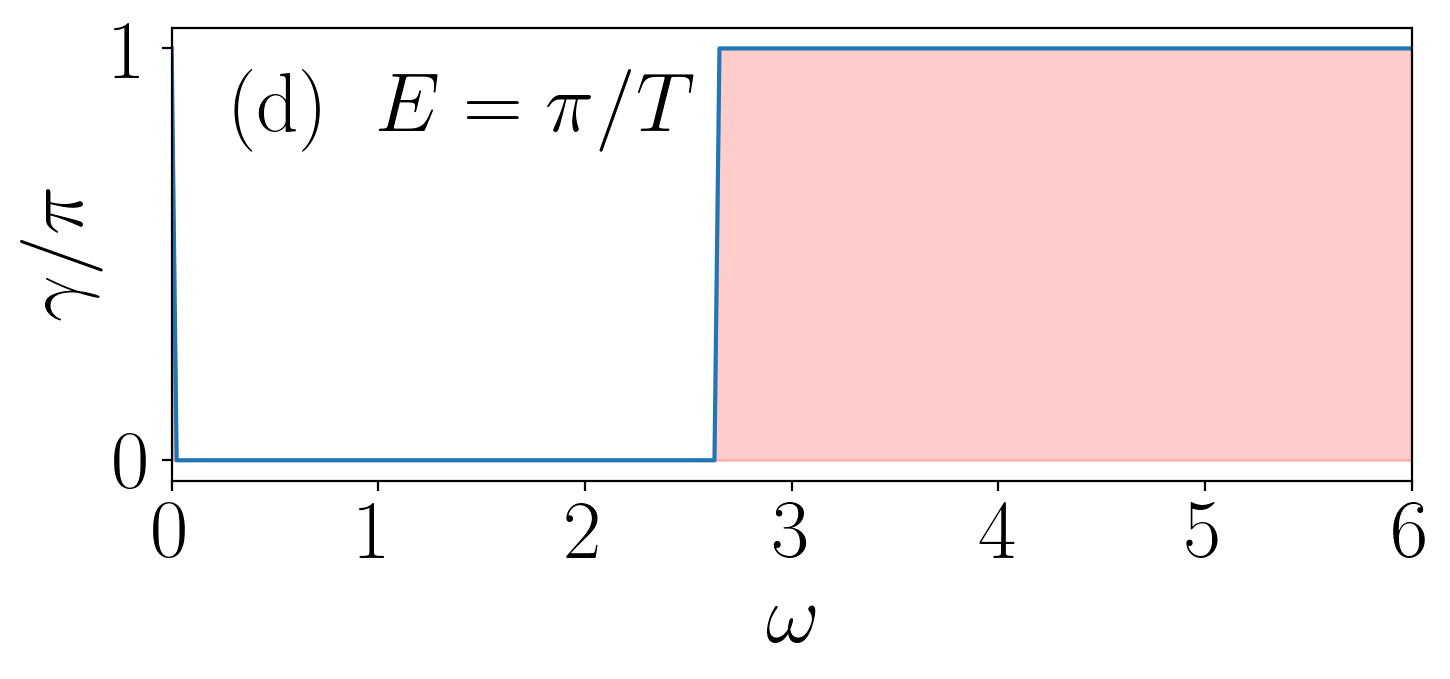}
         \captionlistentry{}
         \label{4.4}
     \end{subfigure}
\caption{{The figures depict total Berry phases for the states filled upto $E = 0$ and $\pi/T$, as a function of the driving frequency, $\omega$. Panel (a) and (c) give the total Berry phases for the zero and the $\pi$ energy modes respectively, under static non trivial condition, with the parameter choice being, $(t_V, t_D) = (1, 1)$. Panel (b) and (d) give the total Berry phases for the zero and the $\pi$ energy modes respectively, under static trivial condition, with the parameter choice being, $(t_V, t_D) = (2.2, 1)$. Other parameters are chosen as, $t_H=1,\theta = \pi/2, V_0 = 0.5$. }} 
 \label{4}
\end{figure}
\end{minipage}
\end{widetext}
\par Once the time dependent perturbation is switched on as shown in Fig.\ref{2.3} and Fig.\ref{2.4} the degeneracies at $\pm \frac{\omega}{2}$ (involving states with $m=0,m=\pm1$) are lifted, leading to the formation of drive induced band gaps, with magnitude  $\Delta\approx 4|V_0|$. One can also verify since the energy spectrum is symmetric about $\pm m\omega/2$, the $\pi$ gap opening relies on the presence of the chiral symmetry of the system. Even when the frequency is very high, the zero and the $\pi$ gaps remain open, but different replicas are widely separated from each other. This prohibits any overlap between the $m=0$, $m\neq 0$ bands. On the other hand, for low frequencies, more replicas start overlapping, and inside the spectrum of $m=0$ we observe large number of degeneracies due to the mixing of different bands.\\

\par In order to show evidence of topological phase transition in our driven scenario, we need to consider the Floquet spectrum of a semi-infinite system in a wide range of frequencies. Due to the existence of an edge, one can infer on the topological character from the presence of edge states. Fig.\ref{3} shows the driven quasi-energy spectrum in both the situations that correspond to trivial ($\frac{t_V}{2t_D}>1$) and topological ($\frac{t_V}{2t_D}<1$) pertaining to the static case. In the subsequent discussions, the parameters such as, $E,t_V,t_D,\omega,V_0$ are measured in units of $t_H$, where, we have set $t_H=1$. As soon as the time dependent perturbation is switched on, the $\pi$ energy modes appear, which have an extended nature upto a certain range of frequencies depending upon the magnitude of the band gap between $ E =\pm \frac{\omega}{2}$ states. However in the limit $\frac{t_V}{2t_D}>1$, for which the system was completely trivial, now in presence of driving shows non-trivial behaviour with the emergence of zero energy modes (See Fig.\ref{3.2}). One can also verify, for the rungless case $(t_V = 0)$, due to flat band like dispersion, there is no possible overlapping between different replicas, and hence, we can not generate localised $\pi$ energy modes even when the drive is intact ($V_0\neq 0$).\\

\par If one can interpret the hopping parameters as $t_V=t_0+\delta$ and $2 t_D=t_0-\delta$, then
analytically we can set a threshold value of the frequency, say $\omega=2t_0$. Above this value, in the static topological limit $(\frac{t_V}{2t_D}<1)$, the system hosts zero energy mode. Whereas, in the static trivial limit $(\frac{t_V}{2t_D}>1)$ the system preserves its zero energy mode upto this threshold frequency. For example in Fig.\ref{3.1}, for the parametric choice $(t_V, t_D) =(1, 1)$, the zero energy mode is formed at a frequency, $\omega$ such that $\omega>2t_0$ ($\omega = 3$ in Fig.\ref{3.1}). Whereas, in Fig.\ref{3.2} for the parametric choice $(t_V,t_D)=(2.2,1.0)$, the zero energy mode is preserved upto a frequency, $\omega$, such that $\omega<2t_0$ ($\omega = 4.2$ in Fig.\ref{3.2}). Hence, we say for frequency, $\omega >2t_0$, close to $E=0$, the topology of the driven system corresponds to that of the undriven one. We label this regime as high frequency regime. Similarly close to $E=\pm\frac{\omega}{2}$, the high frequency regime corresponds to $\omega >4t_0$, where the spectral gaps at $E = \pm \frac{\omega}{2}$  are non existent. One can also notice that in the static non-trivial limit $(\frac{t_V}{2t_D}<1)$, there are no zero energy modes upto $\omega=2t_0$. This can alternatively be verified from the bulk spectrum, where within the frequency range, $\omega<2t_0$, although the spectral gaps at the edges of the BZ are open, there are other gapless points correspond to the $m=0$ replica. As we increase more branches, for example $m=\pm1$ (or even $m=\pm 2$ etc.), spectral gaps open up at these degenrate points. However, these are $\pi$ energy modes, and we have no zero energy modes for $\omega<2t_0$. Based on the above discussions, we may conclude that under static non-trivial condition $(\frac{t_V}{2t_D}<1)$, the driven system has no zero energy mode in the low frequency regime. Whereas, under static trivial condition $(\frac{t_V}{2t_D}>1)$, the driven system has no zero energy mode in the high frequency regime.\\

To further confirm the topological signatures in terms of `bulk-edge correspondence', we resort to the calculation of the topological invariant \cite{Ryu}. The relevant invariant for a $3 \times 3 $ Floquet-Bloch Hamiltonian in the frequency domain is the Berry phase\cite{resta,xiaoberry}, which denotes the geometric phase acquired by a wave function as the system is smoothly taken across the Brillouin zone. A Hamiltonian with a non-trivial Berry phase can not be adiabatically connected to an atomic insulator unless a gap closing transition occurs. The Berry phase is defined as, 
\begin{equation}
\gamma=i\oint dk \langle{u_k}| \nabla_{k}{u_k}\rangle,
\end{equation}
where $\ket{u_k}$ are the Bloch states. The numerical calculation of the Berry phases $\gamma_\alpha$ ($\alpha$ denotes band index, marked with the letters $A-F$ in Fig.(\ref{2.4})) for each of the bands for a particular frequency, say, $\omega=4.5$, in units of $t_H$, is obtained as,
\begin{equation}
\gamma_{\alpha}=
\begin{cases}
0 \quad (\alpha= A, F) \\

\pi \quad (\alpha= B,C,D,E)\\
\end{cases}\quad
\textrm{for} ~
 \Big(\frac{t_V}{2t_D}\Big)>1,
\end{equation}
\begin{equation}
\gamma_{\alpha}=
\begin{cases}
\pi \quad (\alpha= A, F) \\

0 \quad (\alpha= B,C,D,E)\\
\end{cases}\quad
\textrm{for} ~~ 
 \Big(\frac{t_V}{2t_D}\Big)<1.
\end{equation}
It is interesting to note that there is atleast one band below the Fermi level that correspond to non-zero Berry phase, and thus signals that the system is in a topologically non-trivial state. Hence, irrespective of the choice whether $t_V>2t_D$ or otherwise, the topological properties of a driven Hamiltonian is now controlled by frequency of the driving field. One can also verify that the cumulative sum of Berry phases, that is $\gamma=|\textrm{mod}(\sum_{\alpha} \gamma_\alpha,2)|$ corresponds to the edge states found in the real space spectrum. Figs.\ref{4.1} and \ref{4.2} and Figs.\ref{4.3} and \ref{4.4} show the results for the cumulative sum of the Berry phase upto $E=0$ (zero mode) and $E=\frac{\pi}{T}$ ($\pi$ mode) corresponding to the non-trivial and trivial limits for the undriven case.
\vspace*{0.77cm}
\begin{center}{\subsection{\label{sec:level3.2}Floquet-Magnus effective Hamiltonian}}\end{center}
\vspace*{-0.5cm}
\par To analyse the correct analytical  behaviour of the Floquet Hamiltonian for large frequencies, that is $\omega>2t_0$, we resort to the high frequency calculation, which involves a rotating frame transformation in the Floquet formalism. In the rotating frame, given by unitary transformation $S(t)$, the transformed Floquet Hamiltonian Eq.\ref{E6} takes the form,
\begin{equation}
\begin{split}
\tilde{H}_{k}(t) & = S^{\dagger}(t)H_F(t)S(t) \\ & = S^{\dagger}(t)H_k(t)S(t)-iS^{\dagger}(t)\dot{S}(t).
\end{split}
\end{equation}
We may choose to work with a particular choice of rotating frame where the unitary transformation is defined as,
\begin{equation}
    S(t)=e^{i\theta(t)\sigma_x}; \quad \theta(t)=\frac{2V_{0}\sin{\omega t}}{\omega},
\end{equation}
so that the transformed Hamiltonian, $\tilde{H}_{k}(t)$ can be written as,
\begin{equation}
\begin{split}
   \tilde{H}_{k}(t) & =  2t_{H} (\sin{k}) e^{2i\theta\sigma^{+}}  +2t_{H} (\sin{k}) e^{-2i\theta\sigma^{-}}  \\ & +(t_V+2t_D\cos{k})\sigma_x,
\end{split}
\end{equation}
where,
\begin{equation}
  \sigma^{\pm}=\frac{\sigma_z\pm i \sigma_y}{2} . 
\end{equation}
Note that, the modified Hamiltonian ($\tilde{H}_{k}(t)$) shares the same periodicity as that of the original one ($H(t)$). With $\tilde{H}_{k}(t)$ being periodic in time, the Fourier decomposition, $\tilde{H}_{k}(t)=\sum_{p} e^{ip\omega t}\tilde{H}_{p,k}$, with $p=0,\pm1,\pm2,..$ and so on, allows us to write an expansion in powers of the inverse of the driving frequency, namely, $1/\omega$, which is known as the Magnus expansion \cite{eckardt,wangmultiple,benitopwave}. This yields an effective Hamiltonian, $H_{\text{eff}}$ given by,
\begin{equation}
\begin{split}
    H_{\text{eff}} & = \tilde{H}_{0,k}+\frac{1}{\omega}[\tilde{H}_{0,k},\tilde{H}_{{1,k}}]-\frac{1}{\omega}[\tilde{H}_{0,k},\tilde{H}_{-1,k}] \\ & -\frac{1}{\omega}[\tilde{H}_{-1,k},\tilde{H}_{1,k}] + \mathcal{O}\Big(\frac{1}{\omega^2}\Big),
\end{split}
\end{equation}
where, $[\tilde{H}_{p,k},\tilde{H}_{{p^{\prime},k}}]$ denotes a commutator. The convergence criteria of the expansion is given by,
\begin{equation}
    \int_{0}^{T} ||\tilde{H}_{k}(t)dt||<\pi.
\end{equation} 
In our driving scenario, this is equivalent to $\omega>t_V+2t_D$ or equivalently $\omega>2t_0$.
By using the expansion \cite{dattoli1,dattoli2},
\begin{equation}
e^{iz\sin{\theta}}=\sum_{n=-\infty}^{\infty}\mathcal{J}_n(z)e^{in\theta},
\end{equation}
with $\mathcal{J}_n$ being the $n^{th}$ order Bessel function of the first kind, the Fourier component of the transformed Hamiltonian can be written as,
\begin{equation}
\begin{split}
\tilde{H}_{p,k} & = [t_V+2t_D\cos{k}]\sigma_x + [2t_{H}\sin{k}\mathcal{J}_{-p}(\frac{2V_0}{\omega})]\sigma^{+} \\& +[2t_{H}\sin{k} \mathcal{J}_p(\frac{2V_0}{\omega})]\sigma^{-}.
\end{split}
\end{equation}
\begin{figure}[!h]
    \begin{subfigure}[b]{\columnwidth}
         \includegraphics[width=\columnwidth]{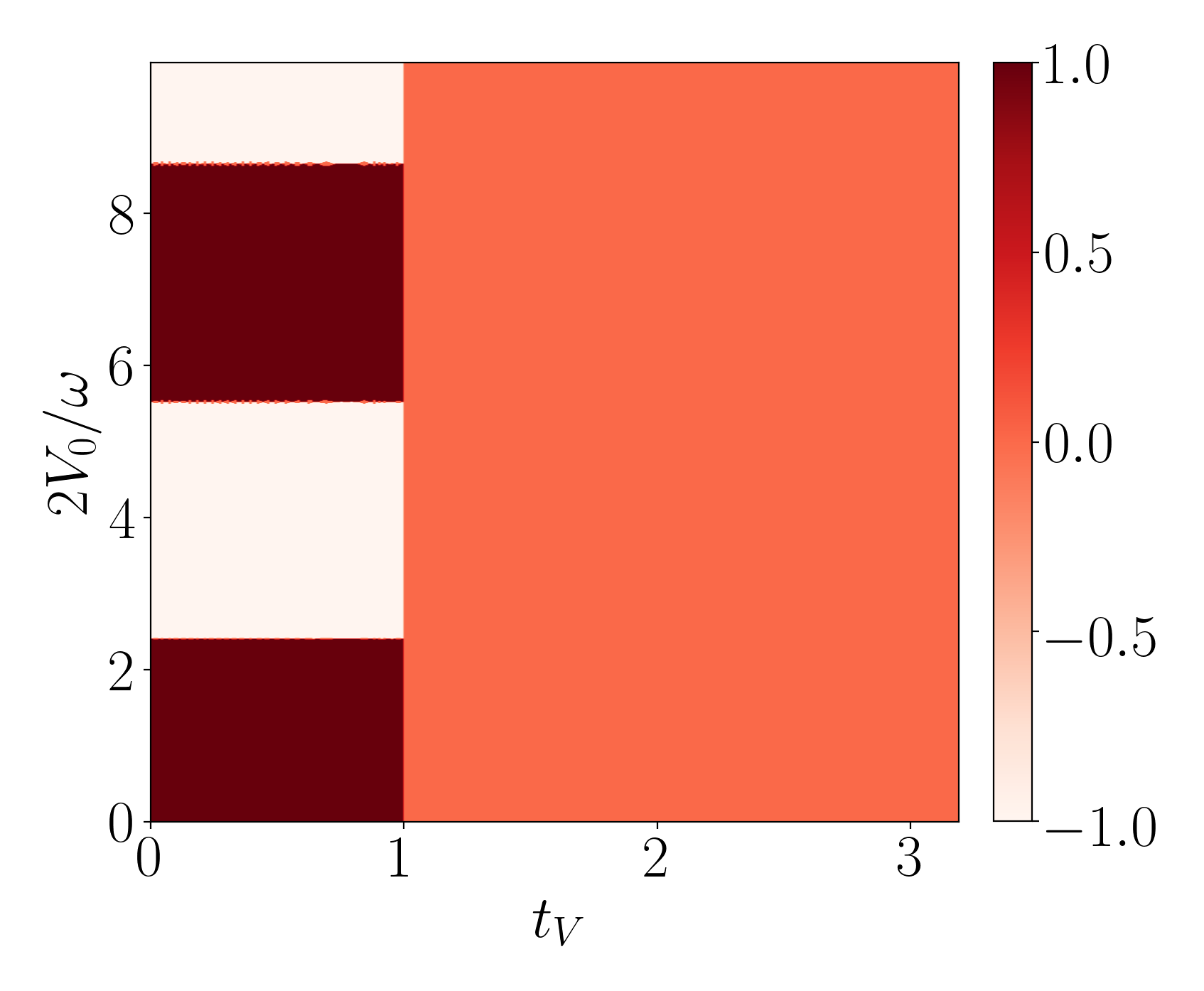}
         \label{5}
     \end{subfigure}
\caption{{The topological phase diagram of the Floquet-Magnus effective Hamiltonian, in the frequency range $\omega >t_V + 2t_D$. The trivial-topological phase transition occurs at $t_V=2t_D$. The parameters used are, $t_D=t_H=0.5,\theta=\pi/2$.}} 
\label{5}
\end{figure}
Within the region of convergence, $\tilde{H}_{0,k}$ is the dominant term, whereas the other components can be neglected owing to the fact that the Bessel functions, $\mathcal{J}_p$ decay rapidly for $p\neq 0$. Hence, the time independent effective Hamiltonian can be written as,
\begin{equation}
H_{\text{eff}}=[t_V+2t_D\cos{k}]\sigma_x+2t_{H}^{\text{eff}}\sin{k}\sigma_{z},
\end{equation}
where, 
\begin{equation}
    t_{H}^{\text{eff}} = t_{H} \mathcal{J}_{0}(\frac{2V_0}{\omega}).
\end{equation}
The effective Hamiltonian in the momentum space can be expressed as a massless Dirac equation of the form,
\begin{equation}
 H(k) = \Vec{d}(k) \cdot \vec{\sigma} , 
\end{equation}
where, $\vec{\sigma}$ denotes the pauli matrices ($\sigma_{x}, \sigma_{y}, \sigma_{z}$) and $d(k)$s are the corresponding vector components, having the forms,
\begin{subequations}
\begin{align}
    d_{x}(k) = t_V+2t_D\cos{k},
    \\ d_{z}(k) = t_{H} \mathcal{J}_{0}(\frac{2V_0}{\omega}) \sin{k}.
\end{align}
\end{subequations}
Hence, the topological properties of the system can be quantified by the invariant called winding number defined as\cite{Ryu,schnyder},
\begin{equation}
\nu=\frac{1}{2\pi}\int_{0}^{2\pi}\frac{d_{z}d(d_x)-d_xd(d_z)}{d_x^2+d_z^2}dk.
\end{equation}
Mathematically, the winding number quantifies how many times the vector $\vec{d}(k)$ winds around the origin as $k$ is varied over the BZ. Physically, the winding number carries the same information as the Berry phase which we have already obtained earlier.\\
\begin{widetext}
    \begin{minipage}{\linewidth}
        \begin{figure}[H]
        \hspace{-12mm}
          \begin{subfigure}[!h]{0.35\columnwidth}
         \includegraphics[width=\columnwidth]{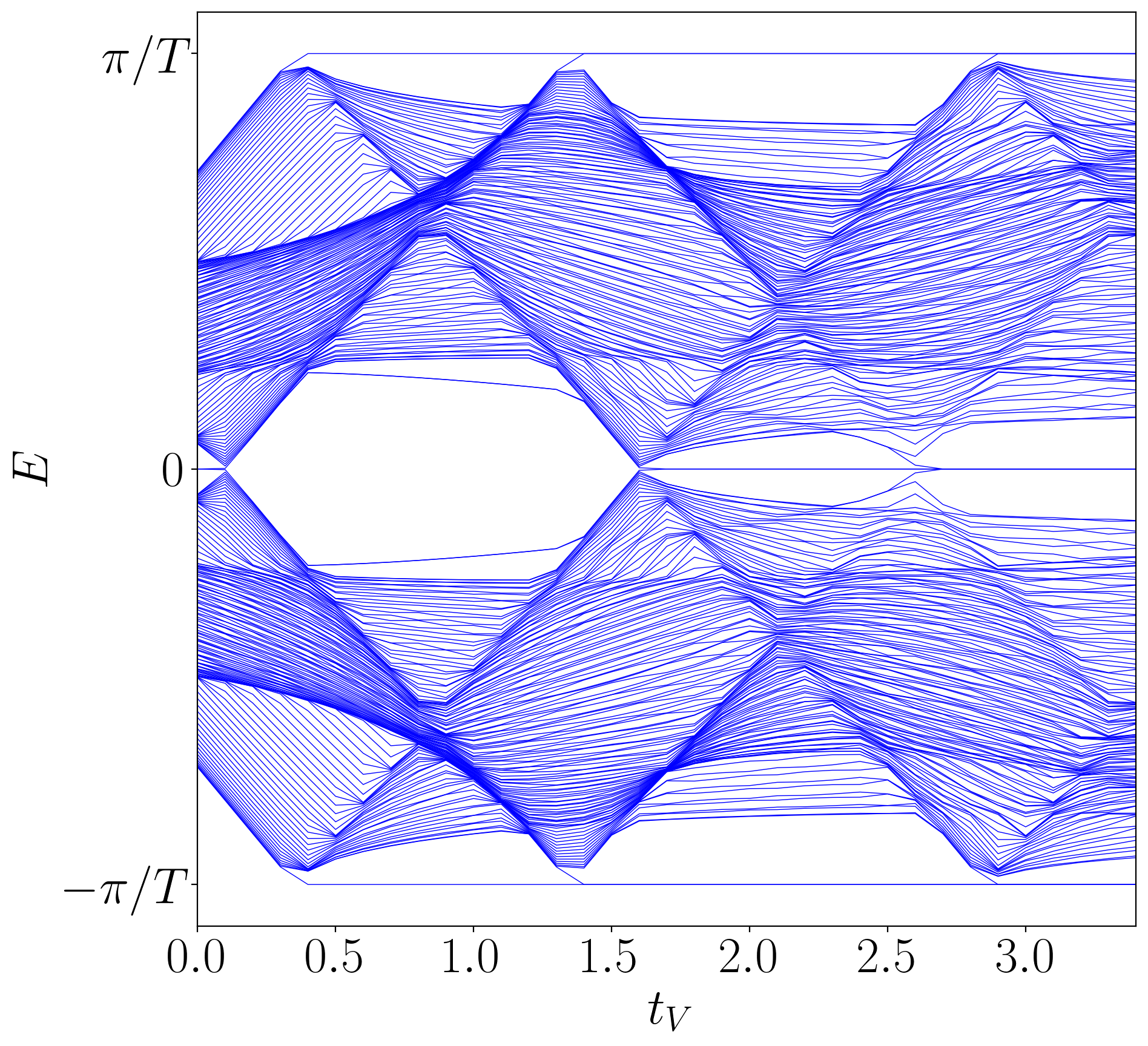}
         \caption{}
         \label{6.1}
     \end{subfigure}
     \begin{subfigure}[!h]{0.35\columnwidth}
         \includegraphics[width=\columnwidth]{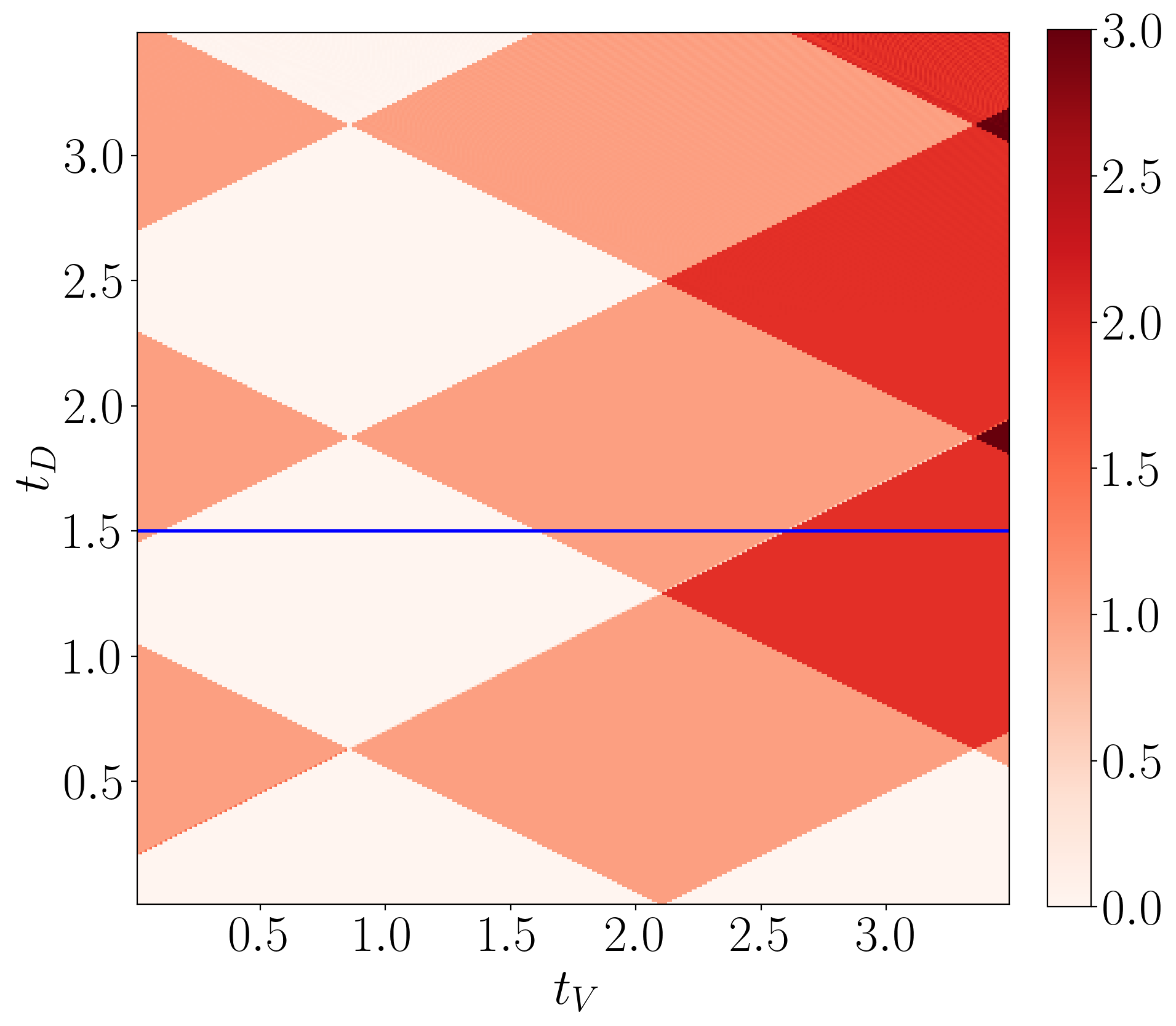}
         \caption{}
         \label{6.2}
     \end{subfigure}
     \begin{subfigure}[!h]{0.35\columnwidth}
         \includegraphics[width=\columnwidth]{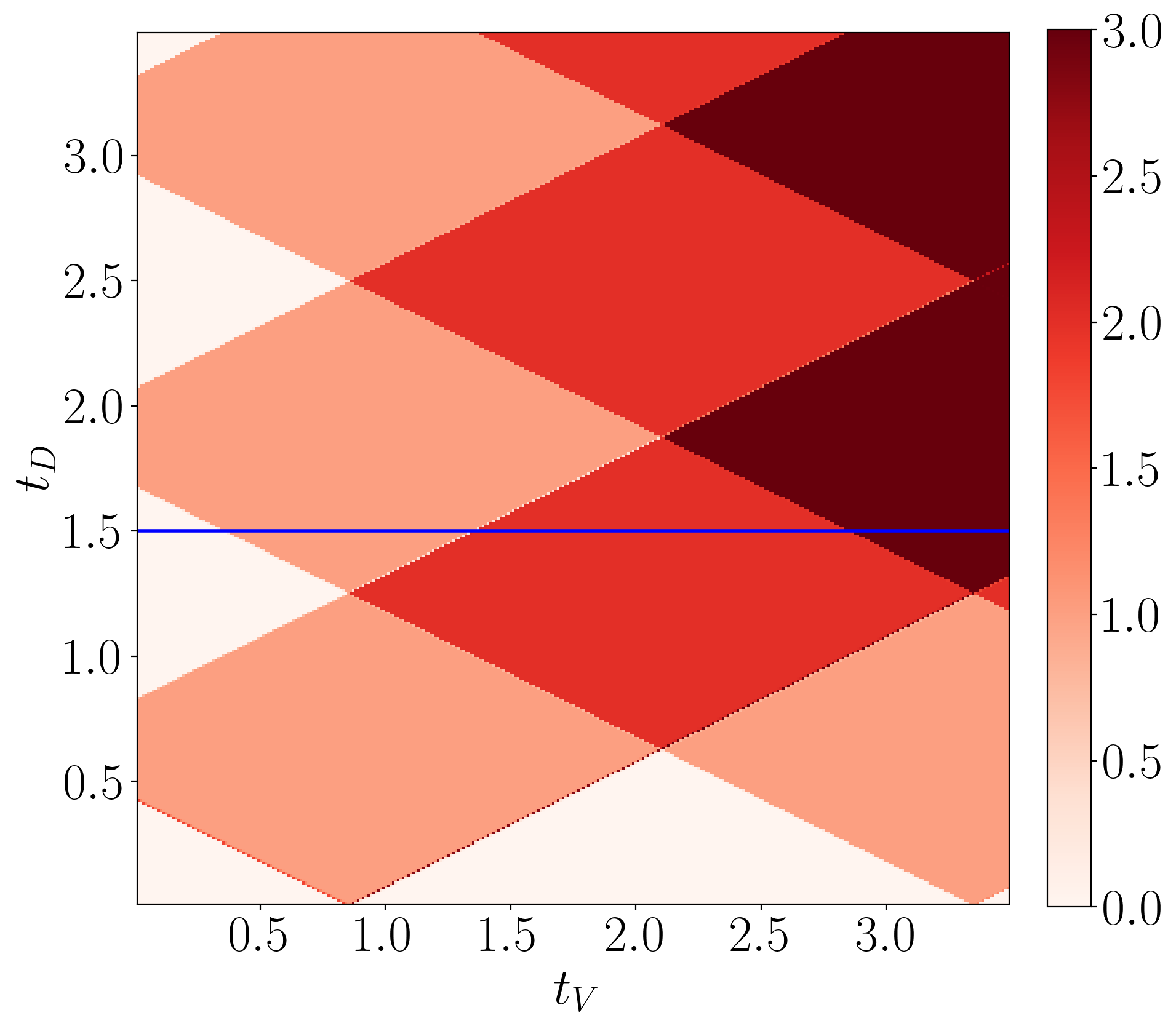}
         \caption{}
         \label{6.3}
     \end{subfigure}
\caption{{(a) shows the Floquet quasi-energy spectrum inside the first FBZ, as a function of the vertical hopping parameter ($t_V$) and with fixed diagonal hopping ($t_D=1.5$) for the $\delta$-kick case. The topological phase transitions are marked by the gap closing at $E = 0$ and $\pi / T$ with the corresponding appearance or disappearance of the zero and $\pi$ energy modes. (b)-(c) depict the topological phase diagram in the $t_V-t_D$ plane, computed using the winding number corresponding to the zero ($\nu^0$) and the $\pi$ ($\nu^{\pi}$) energy modes respectively. The parameters used are, $\omega = 2.5, V_0 = 0.5, t_H=1 $.}} 
 \label{6}
\end{figure}    
\end{minipage}
\end{widetext}
\par From the topological phase diagram plotted in Fig.\ref{5}, we notice that a topological phase transition occurs at $t_V=2t_D$, where the winding number changes from $\nu=0$ to a finite value of $\nu$. Additionally, in the non-trivial region, a new topological phase, characterised by $\nu=-1$ appears. The transition between different topological phases, corresponding to $\nu=1$ and $\nu=-1$ occurs at zeros of the Bessel function, $\mathcal{J}_{0}(\frac{2V_0}{\omega})$. Hence, the different topological phases are characterised by the winding number, $\nu=\textrm{sgn}\mathcal{J}_0 (\frac{2V_0}{\omega})$.
\begin{center}{\subsection{\label{sec:level3.3}$\boldmath{\delta}$-kick}}\end{center}
\par In this section we will discuss another variant of periodic drive, namely, a $\delta$-kick, again associated with the vertical hopping ($t_V$) such that the Hamiltonian becomes, 
\begin{equation}
H_V(t)=\Big[ t_V + V_0 \sum_{m=-\infty}^{m=\infty} \delta(t-mT) \Big] \sum_{n} (a_n^{\dagger}b_n+a_nb_n^{\dagger}).
\end{equation}\\
Unlike the sinusoidal case, the Floquet Hamiltonian corresponding to such a kick can not be expanded in
the frequency domain (See Eq.\ref{E6}). Since the truncation of the infinite dimensional matrix is not possible. Rather, a different technique is followed here. We consider the Floquet time evolution operator defined as,
\begin{equation}
U(T)=\mathcal{T} e^{-i\int_{0}^{T} dt H(t)},
\end{equation}
where $\mathcal{T}$ is the time ordering operator. Now using the Suzuki-Trotter decomposition of first kind \cite{suzuki}, for a $\delta$-driven Hamiltonian, the Floquet time evolution operator  can be written as a product of two exponential matrices, which are,
\begin{equation}
\label{E25}
\begin{split}
    U(T)&= e^{-iV_{0}\sum_{n} (a_n^{\dagger}b_n+a_nb_n^{\dagger})} e^{-iH_0T}  \\ & =e^{-iH_{\text{eff}}T},
\end{split}
\end{equation}
where, $H_{\text{eff}}$ is the time independent effective Hamiltonian analogous to $H_{F}$ (Eq.\ref{E6}) that we had obtained corresponding to the sinusoidal case.
Now to study the Floquet quasi-energy spectrum, we plot the eigenvalues of $H_{\text{eff}}$ which will be confined within the FBZ, namely, ($-\frac{\pi}{T}:\frac{\pi}{T})$ shown in Fig.\ref{6.1}. Similarly in the momentum space Eq.\ref{E25} can be written as,
\begin{equation}
\label{E26}
    \begin{split}
    U_k(T)&=e^{-iV_{0}\sigma_x} e^{-iH_0(k)T}
    \\ & =e^{-iH_{\text{eff}}(k)T}.
\end{split}
\end{equation}
Here, the topological invariant is again winding number. In this scenario, we shall use a slightly different form for the winding number, that is expressed as,
\begin{equation}
\nu[h]=\frac{1}{2\pi i}\int_{BZ} dk \frac{d}{dk} \log{h(k)}.
\end{equation}
To compute the winding number we shall follow the method sketched in Ref.\cite{asbothwinding1}. The chiral symmetry for a generic Hamiltonian of a periodically driven system is written as, $CH(k,t)C^{-1}=-H(k,T-t)$. This ensures that there is an intermediate time $t=\frac{T}{2}$ that splits the period into two parts in a special way \cite{roy10fold}. Let $F$ and $G$ denote the time evolution of the first and the second parts of the cycle respectively. The expressions for $F$ and $G$ are,
\begin{equation}
F=\mathcal{T} e^{-i\int_{0}^{\frac{T}{2}}} H(t)dt,
\end{equation}
\begin{equation}
G=C F^{\dagger}C=\mathcal{T} e^{-i\int_{\frac{T}{2}}^{T} H(t)dt}.
\end{equation}
Hence, using the chiral symmetry, the two topological invariants can be found \cite{asbothwinding1,asbothwinding2} from the half-period evolution operator,
\begin{equation}
U_k  \Big(\frac{T}{2}\Big) =
\begin{bmatrix}
P^{\dagger}_k& Q_k \\
- Q^{\dagger}_k & P_k
\end{bmatrix},
\end{equation}
as, 
\begin{equation}
\label{E30}
    \nu^0=\nu[Q_k]; \quad   \nu^{\pi}=\nu[P_k].
\end{equation}
In the static case, ($V_0=0$) $P_k$ is constant and $Q_k \propto |\vec{d}(k)|$. Thus one finds $\nu^{\pi}=0$ and $\nu^0=\nu$, which is expected.\\

\par Fig.\ref{6.2} and Fig.\ref{6.3} show the topological phase diagrams in the $t_V-t_D$ plane, plotted for certain values of $\omega$ and $V_0$ given by, $\omega=2.5$ and $V_0=0.5$. The confirmation of the bulk-edge correspondence is done by comparing the results with the real space quasi-energy spectrum as a function of $t_V$ plotted for a fixed value of $t_D$, say, $t_D=1.5 $ (See Fig.\ref{6.1}). Note that there are instances of gap closing, not associated with the topological phase transition (as is well known for the static case) but corresponds to higher values of the winding number. One can also verify that as the value of the winding number increases, more number of states from the bulk become localised states and appear at the edges.\\
\begin{figure}[!h]
          \begin{subfigure}[!h]{0.493\columnwidth}
         \includegraphics[height=45mm,width=\columnwidth]{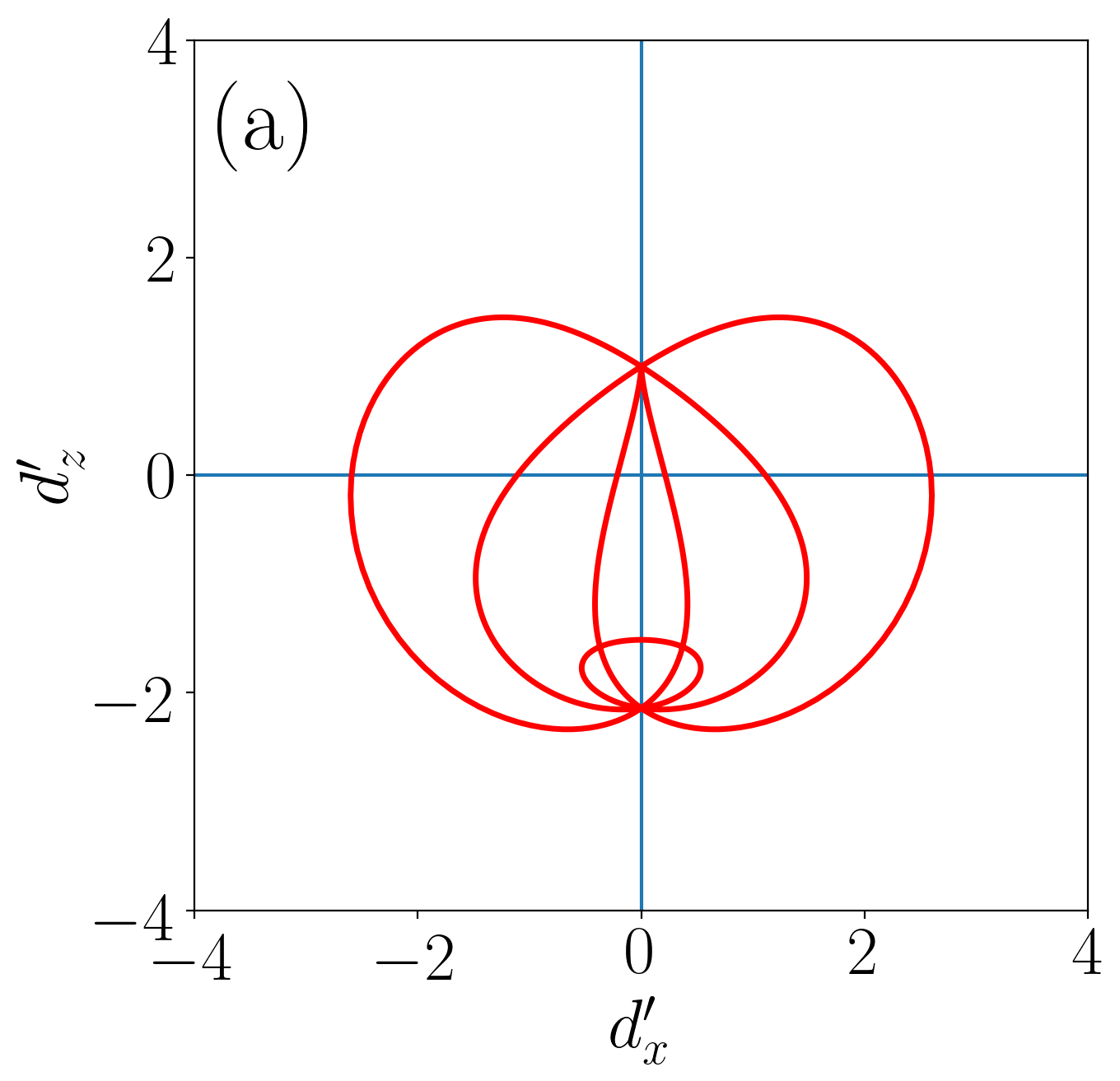}
         \captionlistentry{}
         \label{7.1}
     \end{subfigure}
     \begin{subfigure}[!h]{0.493\columnwidth}
         \includegraphics[height=45mm,width=\columnwidth]{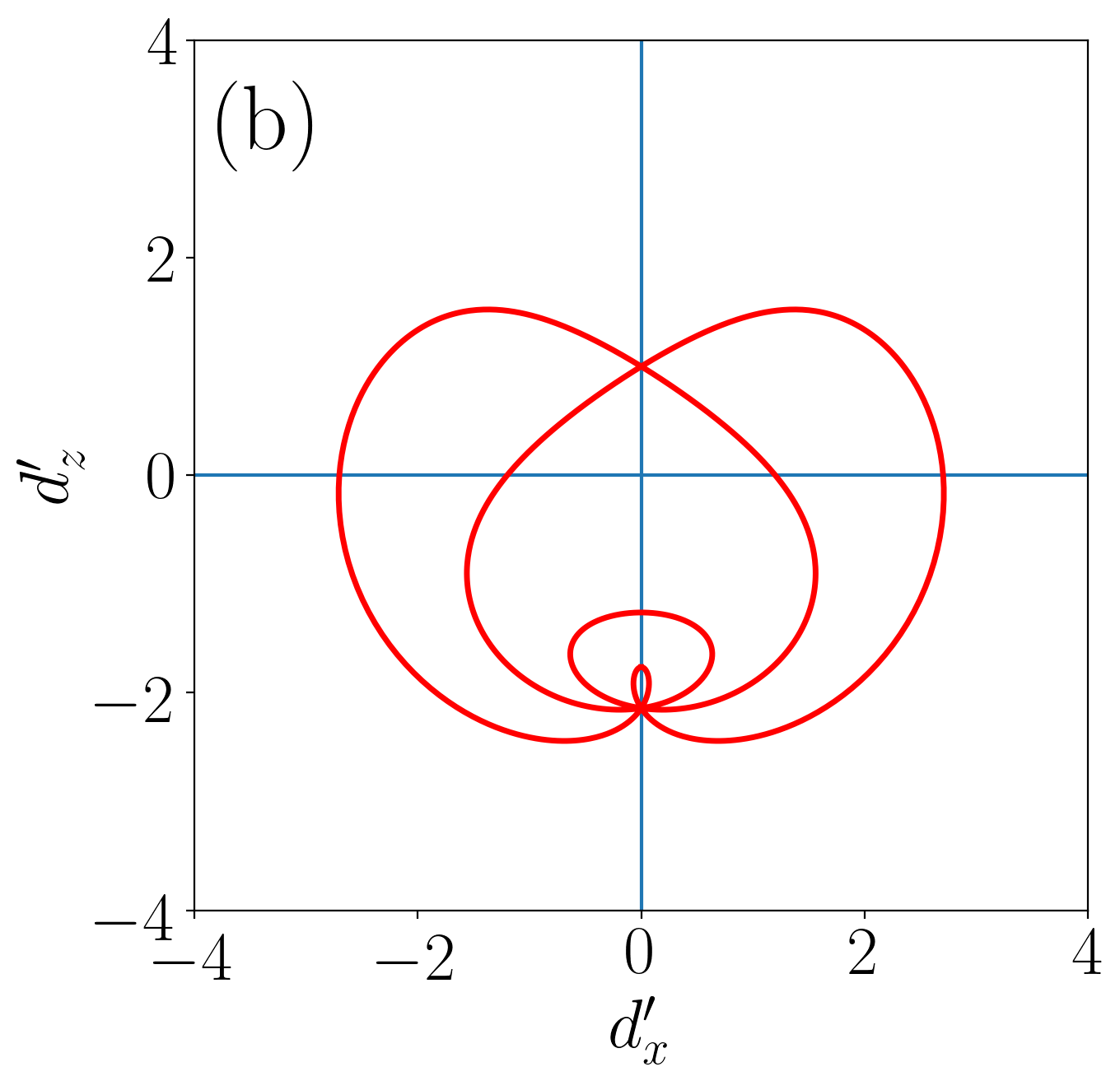}
         \captionlistentry{}
         \label{7.2}
     \end{subfigure}
     \begin{subfigure}[!h]{0.493\columnwidth}
         \includegraphics[height=45mm,width=\columnwidth]{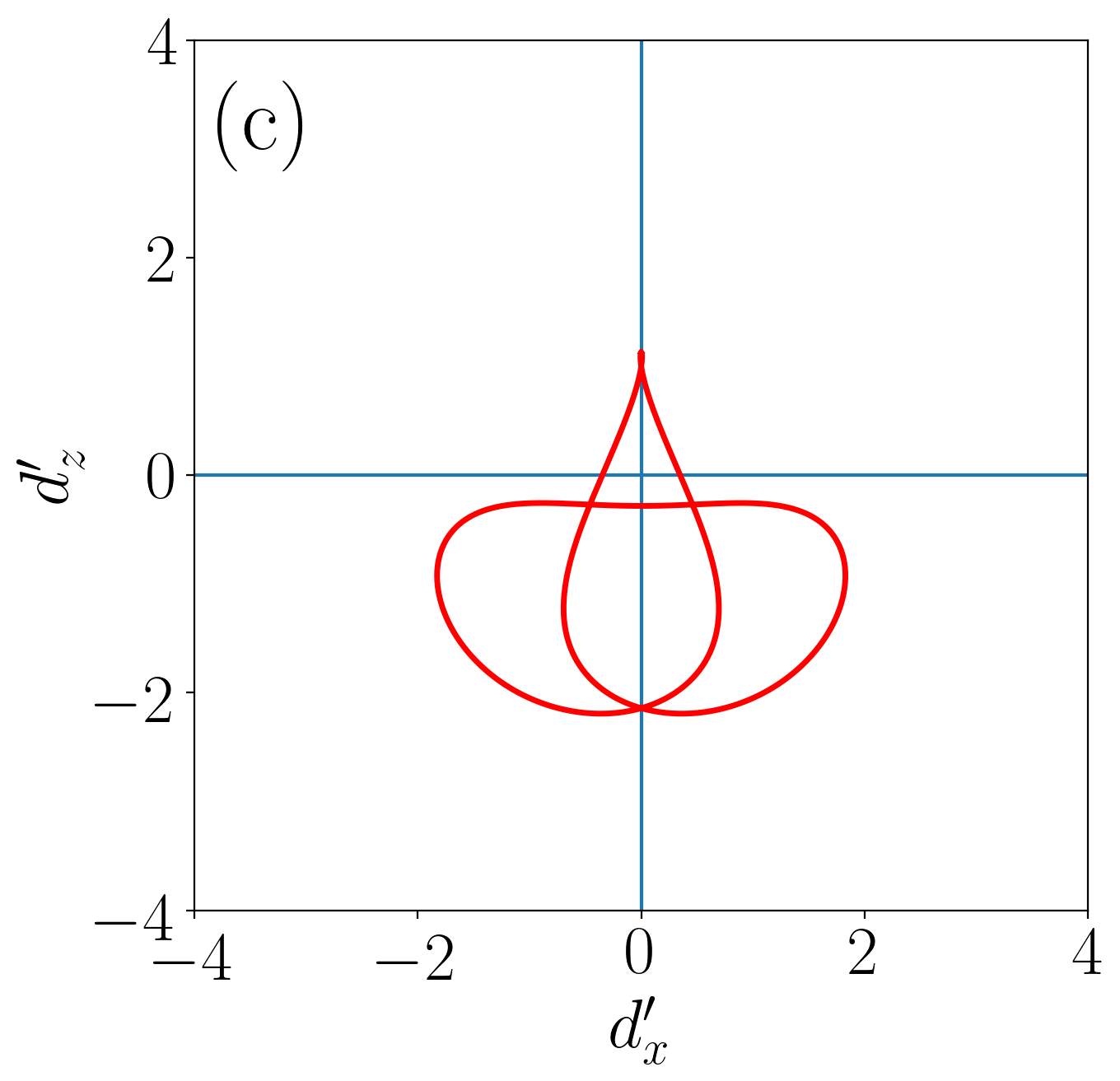}
         \captionlistentry{}
         \label{7.3}
     \end{subfigure}
     \begin{subfigure}[!h]{0.493\columnwidth}
         \includegraphics[height=45mm,width=\columnwidth]{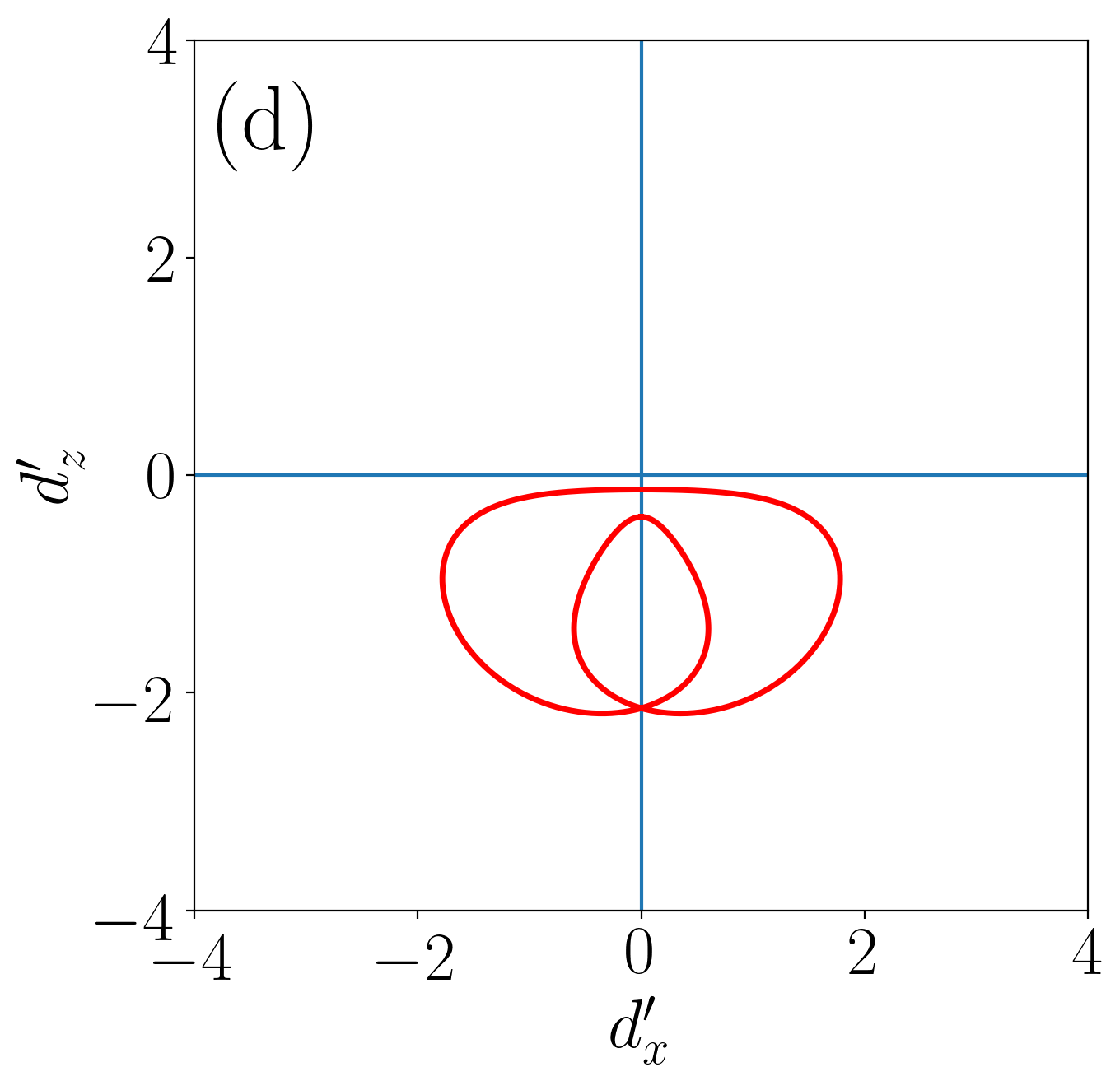}
         \captionlistentry{}
         \label{7.4}
     \end{subfigure}
\caption{{The closed curves in the $d_{x}^{\prime} - d_{z}^{\prime}$ plane for a $\delta$-kick model with fixed $t_H = 1, V_0 = 0.5, \theta = \pi/2$. The parameters used are, ($t_V, t_D) = (2, 1.5)$ in panel (a), (c) and ($t_V, t_D) = (1.5, 1.2)$ in panel (b), (d). Also, the frequency of the drive is chosen as, $\omega=2.5$, for panel (a) and (b), and $\omega=5$, for panel (c) and (d). The corresponding dynamical winding numbers, $\nu^D$ are obtained as, $3,2,1,0$ respectively.}} 
 \label{7}
\end{figure}
\par The effective Hamiltonian obtained in this way has the form of a massless Dirac equation with all the three components of $\vec{d}(k)$ vectors being present. The components of the $d$-vectors in this case are,
\begin{equation}
\begin{split}
    d_{x}(k) = & -\frac{E_k}{\sin(E_{k} T)} \Big[ \sin V_0 \cos (E_{k,0}T) \\ & + \frac{t_V + 2t_D \cos k}{E_{k,0}} \cos V_0 \sin(E_{k,0}T) \Big],
\end{split}
\end{equation}
\begin{equation}
\begin{split}
    d_{y}(k) =  \frac{E_k}{\sin(E_{k} T)} \Big[ \frac{t_H \sin k}{E_{k,0}} \sin V_0 \sin(E_{k,0}T) \Big],
\end{split}
\end{equation}
\begin{equation}
\begin{split}
    d_{z}(k) = & -\frac{E_k}{\sin(E_{k} T)} \Big[ \frac{t_H \sin k}{E_{k,0}} \cos V_0 \sin(E_{k,0}T) \Big],
\end{split}
\end{equation}
where, $E_{k,0}$ is the eigenvalue of the undriven model, and $E_k$ is the quasi-energy of the effective Hamiltonian, given by,
\begin{equation}
\begin{split}
    E_k = \frac{1}{T} \arccos & \Big[ \cos V_0 \cos(E_{k,0}T) \\ & - \frac{t_V + 2t_D \cos k}{E_{k,0}} \sin V_0 \sin (E_{k,0}T)\Big].
\end{split}
\end{equation}
Apparently the chiral symmetry of $H_{\text{eff}}$ looses its meaning. Since, unlike the static version of the Creutz ladder, due to the presence of $\sigma_y$ in the Bloch Hamiltonian, $H_{\text{eff}}$, here the parameter vector $\vec{d}(k)$ does not lie on the $x-z$ plane. Consequently, the winding about any arbitrary axis becomes difficult to visualize. To circumvent this difficulty, let us compute another quantity, called as the dynamical winding number \cite{thakurathi,molignini,mondalkitaev}. To facilitate our computation, we need to rewrite Eq.\ref{E26} in a symmetric way, which can be done using Suzuki-Trotter decomposition of the second kind \cite{suzuki}. As a result, the Floquet time evolution operator in the momentum space assumes a form,
\begin{equation}
\begin{split}
    U_k(T)&=e^{-iV_{0}\sigma_x /2} e^{-iH_0(k)T} e^{-iV_{0}\sigma_x/2}
    \\ & =e^{-iH_{\text{eff}}(k)T},
\end{split}
\end{equation}
where,\vspace*{-0.6cm}
\begin{equation}
H_{\text{eff}}(k)={d_x}^{\prime}(k)\cdot {\sigma}_x+{d_z}^{\prime}(k)\cdot {\sigma}_z.
\end{equation}
The winding number obtained using this Hamiltonian is known as the dynamical winding number. It carries the information regarding both the zero and the $\pi$ energy modes. For example, consider a particular set of parameters, say, $(t_V, t_D)=(2, 1.5)$ and $\omega=2.5, V_0=0.5$. From the topological phase diagram, the winding numbers corresponding to the zero ($\nu^0$) and the $\pi$ ($\nu^{\pi}$) modes are obtained as, $\nu^{0}=1$ and $\nu^{\pi}=2$. The corresponding value for the dynamical winding number ($\nu^D$) in the ${d_x}^{\prime}$-${d_z}^{\prime}$ plane, shown in Fig.\ref{7.1} is obtained as, $\nu^D=3$. Similarly Fig.\ref{7.2} denotes $\nu^D=2$, for other parameter values, namely $(t_V, t_D)=(1.5, 1.2)$ and $\omega=2.5, V_0=0.5$. It can also be checked that for all parameter values, $\nu^D = |\nu^0 + \nu^{\pi}|$, where $\nu^0$ and $\nu^{\pi}$ are defined in Eq.(\ref{E30}). Further, the number of edge modes turns out to be $2 \nu^D$, as we have checked from Fig.\ref{6.1}. One can also confirm that at a frequency, given by, $\omega>2t_0$, the dynamical winding number varies between $0$ and $1$, (See Fig.\ref{7.3}, and Fig.\ref{7.4}), depicting a behaviour similar to the static case in the high frequency regime.
\vspace*{1cm}
\begin{center}{\section{\label{sec:level4}Final remarks}}\end{center}
\vspace*{-0.6cm}
\par For the undriven Creutz ladder, the topological and the trivial limits are set by $\frac{t_V}{2t_D}<1$ and $\frac{t_V}{2t_D}>1$ respectively. We have taken these values as benchmarks and explored the scenario in a driven system. For the sinusoidal drive, we get at least one band to retain its topological character in the trivial limit ($\frac{t_V}{2t_D}>1$) as seen from the non-zero Berry phase. Further, the situation nicely distinguishes between the low and the high frequency regimes of the driving potential. Corresponding to the trivial (topological) case, the zero energy mode ceases to exist above (below) a certain value of the driving frequency. The high frequency limit understandably reproduces the results for the static case, which we have explicitly verified using Floquet-Magnus expansion. On the other hand, in presence of $\delta$-kick, the Creutz ladder presents larger values of the topological invariant, namely the dynamical winding number, and as a consequence of that we have large number of edge modes in the system.\\

\par It is important to deliberate whether we can define winding number as the topological invariant for the sinusoidal drive. The answer is affirmative. But unlike the case of the periodic $\delta$-kick, the Floquet operator can no longer be expressed as a product of just two operators. Rather, it has to be computed by dividing the time period $T$ into a large number of time steps, each of width $\Delta t$, where, $\Delta t = \frac{T}{N}$ and $N$ denotes the number of time steps. Hence, one can multiply each of the operators in a time ordered fashion. The corresponding numerical computation consumes much more time, although it can be done. Indeed the corresponding results are verified by us, and are in excellent agreement with those already obtained from the analysis of the Berry phase. 
\begin{center}{\section{\label{sec:level4}Acknowledgement}}\end{center}
\par Shortly before submission of the manuscript we became aware of the references \cite{zhouspinful,zhounonhermitian} where spinful and non-Hermitian Creutz ladder are studied.

\end{document}